\documentclass[sn-basic]{sn-jnl}
\usepackage{cite}
\usepackage{amsmath,amssymb,amsfonts}
\usepackage{algorithmic}
\usepackage{graphicx}
\usepackage{textcomp}
\usepackage{xcolor}
\usepackage{caption}
\usepackage{multicol}
\usepackage{import}
\usepackage{balance}
\usepackage{float}
\usepackage{enumitem}
\usepackage[dvipsnames, table]{xcolor}
\usepackage{array}
\usepackage{tabularx}
\usepackage{xspace}
\usepackage{makecell}
\usepackage{booktabs}
\usepackage{afterpage}
\usepackage{subfig}
\usepackage{hyperref}
\usepackage{cleveref}
\usepackage{adjustbox}
\usepackage{rotating}
\usepackage{mdframed}
\usepackage{listings}
\usepackage{multirow}
\usepackage{xstring}
\usepackage{threeparttable}

\usepackage[utf8]{inputenc} % for → arrow character if not already set

\lstdefinelanguage{JavaScript}{
  keywords={break, case, catch, continue, debugger, delete, do, else, finally, for, function, if, in, instanceof, new, return, switch, this, throw, try, typeof, var, void, while, with, require, module, exports},
  morecomment=[l]{//},
  morecomment=[s]{/*}{*/},
  morestring=[b]',
  morestring=[b]",
  ndkeywords={class, export, boolean, throw, implements, import, this},
  keywordstyle=\color{blue}\bfseries,
  ndkeywordstyle=\color{darkgray}\bfseries,
  identifierstyle=\color{black},
  commentstyle=\color{purple}\ttfamily,
  sensitive=true
}
\definecolor{diffadd}{RGB}{230, 255, 230} 
\definecolor{diffrem}{RGB}{255, 230, 230} 
\definecolor{addtext}{RGB}{0, 100, 0}      
\definecolor{remtext}{RGB}{150, 0, 0}      

\lstdefinelanguage{json}{
    basicstyle=\ttfamily\small,
    columns=fullflexible,
    breaklines=true,
    frame=single,
    showstringspaces=false,
    commentstyle=\color{gray},
    numbers=left,
    numberstyle=\tiny\color{gray},
    numbersep=12pt,        
    xleftmargin=20pt,      
    tabsize=2
}

\newmdenv[
  backgroundcolor=light-gray,
  linecolor=black,
  linewidth=1pt,
  innerleftmargin=8pt,
  innerrightmargin=8pt,
  innertopmargin=8pt,
  innerbottommargin=8pt,
  roundcorner=10pt,
  nobreak=true
]{answerbox}

\definecolor{light-gray}{gray}{0.95}
\definecolor{light-gray-bg}{gray}{0.9}
\definecolor{pgreen}{RGB}{5,205,107}
\definecolor{pblue}{RGB}{2,154,223}

\newcommand{\edit}[1]{\textcolor{black}{ #1}}

\newcommand{\javascript}{JavaScript\xspace}

\newcommand{\npm}{npm\xspace}

\newcommand{\package}{\texttt{package.json}\xspace}

\newcommand{\void}{\texttt{Void}\xspace}
\newcommand{\core}{\texttt{Core}\xspace}
\newcommand{\develop}{\texttt{Dev}\xspace}
\newcommand{\peer}{\texttt{Peer}\xspace}
\newcommand{\arrow}{~$\rightarrow$~}

\newcommand{\RQone}{ How prevalent do developers reclassify dependencies? \xspace}
\newcommand{\RQtwo}{ How do developers remove dependencies in practice? \xspace}
\newcommand{\RQthree}{ How do developers reassign dependency roles in practice? \xspace}
\newcommand{\RQfour}{ How long does it take developers to reclassify dependencies? \xspace}

\newcommand{\totaloriprojects}{ 52,589 }
\newcommand{\nolongerprojects}{ 5,636 }

\newcommand{\totalprojects}{ 33,087 }
\newcommand{\totalonlyadd}{ 6,900 }

\newcommand{\totalcommits}{ 1,504,472 }
\newcommand{\totalreccommits}{ 291,741 }

\begin{document}
\title{A Longitudinal Study of Dependency Reclassifications in \javascript Projects}

\author*[1]{\fnm{Yuxin} \sur{Liu}}\email{yuxinli@kth.se}

\author[1]{\fnm{Cristian} \sur{Bogdan}}\email{cristi@kth.se}

\author[2]{\fnm{Benoit} \sur{Baudry}}\email{benoit.baudry@umontreal.ca}

\affil[1]{\orgname{KTH Royal Institute of Technology}, \country{Sweden}}
\affil[2]{\orgname{Université de Montréal}, \country{Canada}}

\abstract{
Modern software projects depend on third-party dependencies, whose declarations must be maintained as projects evolve.
Prior work has focused on dependency version updates, while much less is known about how developers assign dependencies to different roles over time.
In this paper, we investigate how developers of \javascript projects reclassify their dependencies, including removal and role reassignment.
By analyzing commit-level modifications to \texttt{package.json} files, we reconstruct dependency role histories and identify recurring reclassification practices.
Our analysis of
\totalprojects \javascript projects with active dependency maintenance reveals that dependency reclassification is a prevalent maintenance activity, occurring in 79.1\,\% of the studied projects, and accounting for 19.4\,\% of all dependency-maintenance commits.
Of these projects, nearly all (97.2\,\%) remove dependencies at some point, while 38.0\,\% undergo role reassignments across Core (runtime), Dev (development-only), and Peer (consumer-provided) roles.
These changes are not always final, as 33.1\,\% of projects later reintroduce removed dependencies and 11.2\,\% exhibit repeated role switching.
Reclassification practices also tend to unfold over long periods, with a median duration of 408 days.
Illustrative cases further show recurring motivations: batch removals as accumulated cleanup or toolchain retirement; reassignments from Core to Dev, which revise production install scope for auditing or builds; and Peer reassignments that simplify consumer installation or move packages that consumers need not provide into Dev.
These findings broaden the study of dependency maintenance beyond version updates, and contribute implications for project maintainers, dependency tools, package managers, and research to support correct dependency role declarations.
}

\maketitle

% \clearpage
\section{Introduction}
Modern \javascript projects rely heavily on third-party packages, commonly known as dependencies, to accelerate development by reusing existing functionality \citep{mohagheghi2004empirical}.
Although adding dependencies is straightforward, maintaining them over time is considerably more difficult \citep{jafari2021dependency, mens2024overview}.

A key part of dependency maintenance is deciding the role of a dependency.
In \javascript projects, package managers such as \npm install packages according to the roles declared in \package: Core dependencies (\texttt{dependencies}) are installed for production, Dev dependencies (\texttt{devDependencies}) are used only during development and are omitted from production installs, and Peer dependencies (\texttt{peerDependencies}) are expected to be provided by the consuming project~\citep{npm_declare_dependencies}.
Unlike ecosystems that primarily distribute compiled binaries, \npm installs package contents as source code that the project can load and execute directly.
Consequently, the declared role does not merely label a package in the manifest: it decides which source is fetched into the install tree and which source can enter the runtime environment (Core).
A role change therefore changes the project's install surface, and, through transitive dependencies, can also change what downstream consumers install.

What ends up in that install surface has practical costs.
If an unused package, or a development tool, is declared as Core, production installation grows larger and more code is present at runtime than the project needs~\citep{liu2025detecting}.
Maintainers then have more packages to track and update, and unused dependencies can also waste CI time when they keep triggering builds~\citep{weeraddana2024dependency, jafari2021dependency}.
Beyond these day-to-day effects on performance and maintenance, a larger install surface may increase security exposure~\citep{koishybayev2020mininode}.

When developers revise dependency declarations, they remove the dependency or reassign it to another role. 
These revisions are not limited to a single commit.
Just like code, dependency manifests such as \package require iterative refinement.
As projects evolve, the most appropriate role for a dependency may become clearer, and developers may revisit earlier assignments to better match actual usage.
In this study, we refer to any such modification of a dependency’s declared role after its initial introduction as a \textbf{dependency reclassification}.

Prior research on dependency management has largely treated dependency roles as stable, with most attention given to versioning~\citep{decan2018evolution, kula2018developers, salza2018developers, cogo2019empirical, javan2023dependency, jayasuriya2022towards}, security issues~\citep{liu2022demystifying, jafari2023dependency}, and adoption trends~\citep{kikas2017structure, abdalkareem2017developers, mujahid2023go}.
As a result, dependency maintenance is observed mainly through additions and version updates, while the reorganization of declared roles over project history remains unexplained.

We study this dependency maintenance at scale.
In this study, we conduct the first large-scale longitudinal analysis of dependency reclassification across \totalprojects \javascript projects that actively maintain their dependencies, which together contain \totalcommits dependency-maintenance commits. 
We reconstruct each dependency's declared role history, identify reclassification events (including removals, role reassignments, and subsequent reversions), and characterize recurring practices by their prevalence, structural forms, and timing.

Our study shows that dependency roles are actively managed long after their initial introduction.
Among the studied projects, 79.1\,\% reclassify at least one dependency after its introduction, and the corresponding reclassification commits account for 19.4\% (\totalreccommits/\totalcommits) of all dependency-maintenance commits.
Beyond version updates, developers frequently revise these roles through removal and reassignment. 
For example, Core-to-Dev reclassifications redefine the boundary between runtime and development dependencies, while Core-to-Peer reclassifications shift installation responsibility between projects and their consumers.
These adjustments often span years of project history, showing that role maintenance is a continuous process.
Taken together, these findings broaden the empirical understanding of dependency maintenance in \javascript projects.

Our primary contributions are outlined as follows:
\begin{itemize}[label= -]
    \item We present, to the best of our knowledge, the first large-scale longitudinal study of dependency reclassification in \javascript projects.
    \item We establish an empirical taxonomy of reclassification practices, and analyze their prevalence, structural forms, and timing across projects.
    \item We derive implications for maintainers, tooling, and package managers on how role review, role-history visibility, and clearer dependency installation guidance can support dependency maintenance.
    \item We make our data publicly available to support future research: \\ \href{https://github.com/ASSERT-KTH/dependency-reclassification}{https://github.com/ASSERT-KTH/dependency-reclassification}
\end{itemize}

\section{Background and Motivation}
This section introduces dependency role declarations in \javascript projects and uses an example project to show why their later revision deserves explicit study.

\subsection{\javascript dependency management}
In \javascript projects, dependencies are declared in the \package file,\footnote{See \texttt{package.json} in \href{https://docs.npmjs.com/cli/v10/configuring-npm/package-json}{npm}, \href{https://classic.yarnpkg.com/lang/en/docs/package-json/}{yarn}, and \href{https://pnpm.io/package_json}{pnpm}.} which serves as the main configuration and dependency manifest.
Each dependency is declared to a specific role based on its intended usage context:
\begin{itemize}
    \item Core Dependencies (\texttt{"dependencies"} in the \package file): required at runtime in production.

    \item Dev Dependencies (\texttt{"devDependencies"}): used only for development purposes, such as testing and build processes.
    
    \item Peer Dependencies (\texttt{"peerDependencies"}): required at runtime, but expected to be provided by the consuming projects, common in plugin and library ecosystems. 
\end{itemize}
\bigskip

To illustrate how dependency roles are declared in practice, \autoref{lst:pkgjson-vue-axios} presents an excerpt from the \package file of the open-source project \texttt{vue-axios}.\footnote{\href{https://github.com/imcvampire/vue-axios/tree/aded81c2}{vue-axios} is a lightweight wrapper that integrates the HTTP client \texttt{axios} into the \texttt{Vue.js} UI rendering framework.}
In this version, the project declares Core dependencies such as \texttt{vue} and \texttt{axios}, several Dev dependencies such as \texttt{@babel/core} and \texttt{gulp}, and also lists \texttt{vue} as a Peer dependency.
Peer declarations are less common than Core and Dev.
Their purpose is mainly to name a host package that consumers already provide (e.g., \texttt{vue}), so that plugins and libraries do not install a second copy.
Omitting a needed Peer declaration can leave consumers without a clear install requirement; declaring a host package as Core instead can create duplicate installations and version conflicts.

\begin{lstlisting}[
  caption={Excerpt from the \href{https://github.com/imcvampire/vue-axios/commit/9ffe3eab06937eac1620991a71235322b6a4f1c1}{\texttt{package.json}} file of the project \texttt{vue-axios}} on 2020-10-08, 
  basicstyle=\footnotesize\ttfamily,
  label={lst:pkgjson-vue-axios}
]
{
  "name": "vue-axios",
  "version": "3.0.1",
  "dependencies": {
    "axios": "^0.20.0",
    "semver": "^7.3.2",
    "vue": "^2.0.0"
  },
  "devDependencies": {
    "@babel/core": "^7.11.6",
    "@babel/preset-env": "^7.11.5",
    "gulp": "^4.0.2",
    "gulp-babel": "^8.0.0",
    "gulp-clean": "^0.4.0",
    "gulp-rename": "^2.0.0",
    "gulp-uglify-es": "^2.0.0",
    "gulp-uglifyjs": "^0.6.2"
  },
  "peerDependencies": { "vue": ">= 3.0.0" }
}
\end{lstlisting}

\subsection{Motivation}
\autoref{lst:pkgjson-vue-axios} shows how \texttt{vue-axios} declares Core, Dev, and Peer dependencies at one point in its history.
In practice, such declarations are not fixed: maintainers may later revise whether a package should remain in the production set, stay only for development, or be provided by consumers.
A concrete case from \texttt{vue-axios} illustrates why these later revisions matter.

{In October 2020, a consumer opened issue~\href{https://github.com/imcvampire/vue-axios/issues/72}{\#72} after an \texttt{npm audit} against a production install of \texttt{vue-axios} reported high-severity vulnerabilities.
The findings were attributed through \texttt{vue-axios}'s own Core declarations to \texttt{rollup-babel}, a superseded build helper that had been placed under \texttt{dependencies} together with related bundling tooling.
The reporter argued that such packages belong in \texttt{devDependencies}, because they support packaging the library rather than running it in downstream applications.
Misplacing them under Core therefore enlarges the install surface that consumers pull and that security audits attribute to the package.

\autoref{lst:reclassification-vue-axios} shows the maintainer response merged in pull request~\href{https://github.com/imcvampire/vue-axios/pull/76}{\#76} (commit~\href{https://github.com/imcvampire/vue-axios/commit/c6a1b2f}{\texttt{c6a1b2f}}, 2020-11-16, message: \texttt{correct dependencies}).
The same commit makes three distinct role decisions:
First, the maintainers remove \texttt{rollup-babel} from Core.
This change directly addresses the consumer report by removing the package, together with its transitive vulnerabilities, from installations of \texttt{vue-axios}.

Second, they move build-oriented packages such as \texttt{gulp-file} from Core to Dev.
These packages remain necessary for developing and packaging \texttt{vue-axios}, but are no longer included in its production dependency set.
This revision therefore separates the project's development toolchain from the dependencies required by consumers at runtime.

Third, and separately, they move \texttt{axios} from Core to Peer.
This revision is not required by the audit findings in issue~\href{https://github.com/imcvampire/vue-axios/issues/72}{\#72}; instead, it changes the responsibility for providing \texttt{axios}.
Declaring \texttt{axios} as a Peer dependency expresses that it should be provided in the consuming project's dependency environment rather than maintained as an ordinary production dependency of \texttt{vue-axios}.

\begin{lstlisting}[
  caption={Excerpt from the diff of \href{https://github.com/imcvampire/vue-axios/commit/c6a1b2f}{\texttt{package.json}} file of the project \texttt{vue-axios}} on 2020-11-16, 
  basicstyle=\footnotesize\ttfamily,
  label={lst:reclassification-vue-axios}
]
(*\colorbox{diffrem}{\color{remtext}{- \quad "dependencies": \{}}*)
(*\colorbox{diffadd}{\color{addtext}{+ \quad "devDependencies": \{}}*)
     "axios": "^0.20.0",
     "gulp-file": "^0.4.0",
     // ... 
(*\colorbox{diffrem}{\color{remtext}{- \quad \quad "rollup-babel": "\^{}0.6.3"}}*)
     // ... 
(*\colorbox{diffrem}{\color{remtext}{- \quad \quad "vue": "\^{}2.0.0"}}*)
(*\colorbox{diffrem}{\color{remtext}{- \quad \},}}*)
(*\colorbox{diffrem}{\color{remtext}{- \quad "devDependencies": \{}}*)
(*\colorbox{diffadd}{\color{addtext}{+ \quad \quad "vue": "\^{}3.0.0",}}*)
     // ... 
     "rollup": "^2.32.1"
   },
   "peerDependencies": {
(*\colorbox{diffadd}{\color{addtext}{+ \quad \quad "axios": ">= 0.20.0"}}*)
   }
\end{lstlisting}

\bigskip

The three changes therefore serve distinct purposes: removing a problematic build package and its transitive vulnerabilities, correctly scoping the remaining build tooling as development-only, and shifting responsibility for the runtime library \texttt{axios} to the consuming project.

\begin{figure}
    \centering
    \includegraphics[width=0.95\linewidth]{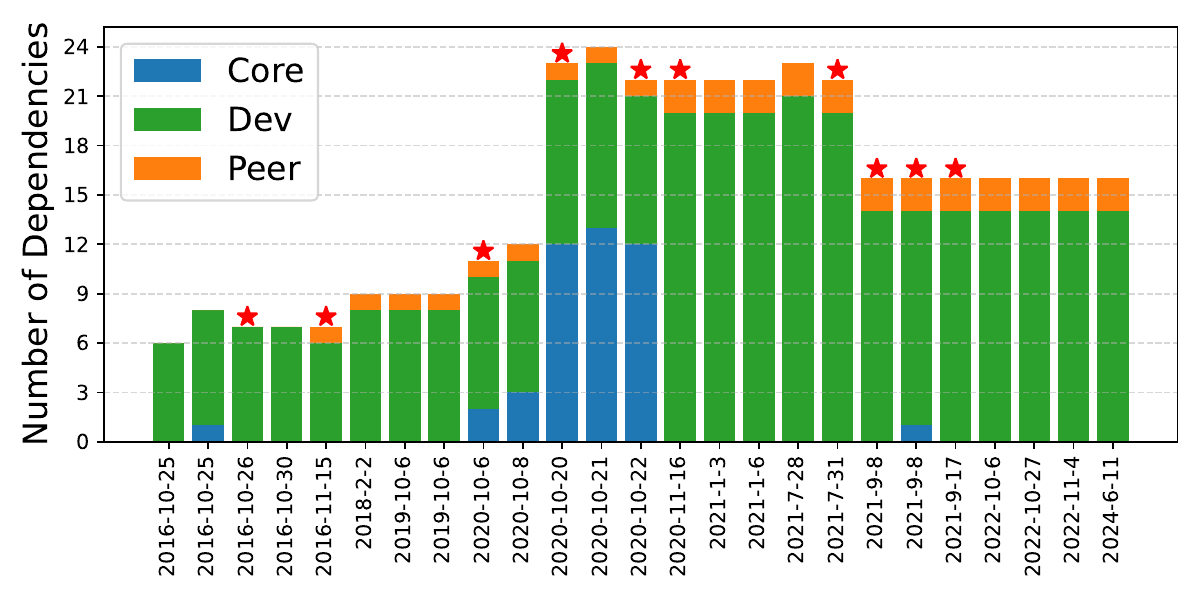}
    \caption{\small
    Dependency declarations in the \texttt{vue-axios} project (2016-10-25 to 2024-06-11).
    Each bar is a dependency-maintenance commit and shows the number of dependencies declared under Core (\texttt{dependencies}), Dev (\texttt{devDependencies}), and Peer (\texttt{peerDependencies}).
    Red stars mark commits where at least one dependency is reclassified.
    The starred commit on 2020-11-16 corresponds to pull request~\href{https://github.com/imcvampire/vue-axios/pull/76}{\#76}, which closes issue~\href{https://github.com/imcvampire/vue-axios/issues/72}{\#72}.
    }
    \label{fig:motivation}
\end{figure}
\bigskip

Such revisions matter because package managers install dependencies according to declared roles, but do not check whether a role matches intended usage.
A development tool declared as Core does not trigger a build error; it can remain in the production set, and in consumer audit reports, until maintainers revise the declaration.
The \texttt{vue-axios} case shows that role declarations are not a purely local manifest concern: they shape both the project's own production dependency set and what consumers actually install and see in their audits.

Beyond individual commits, it is unclear whether such revisions are isolated or recur throughout a project's lifecycle.
To examine this more closely, we further analyze the full history of dependency declarations in the \texttt{vue-axios} project, and visualize the number of dependencies over time in \autoref{fig:motivation}.
Each bar is a dependency-maintenance commit; red stars mark commits with at least one reclassification.
One starred commit is the response to issue~\href{https://github.com/imcvampire/vue-axios/issues/72}{\#72} in pull request~\href{https://github.com/imcvampire/vue-axios/pull/76}{\#76} (2020-11-16); the remaining stars mark earlier and later role revisions in the same project.
The figure shows that Core, Dev, and Peer counts continue to change throughout project history, so reclassification can recur and unfold over extended periods of evolution.

This case illustrates the motivation for studying dependency reclassification.
Prior work largely tracks version updates within a dependency's current role~\citep{javan2023dependency}, or summarizes projects by how many dependencies they declare~\citep{kikas2017structure}.
Neither view captures the revisions above.
Moving build tooling from Core to Dev, or \texttt{axios} from Core to Peer, may leave version strings and even the total number of declared dependencies largely unchanged, yet it changes who installs what.
We therefore analyze reclassification histories at scale to characterize how prevalent these role revisions are, what forms they take, and how long such reclassification processes take across \javascript projects.

\section{Research Method}\label{sec:research-method}
This section describes the research method used to investigate dependency reclassification at corpus scale.
We state the research questions that guide the study (Section~\ref{method:rqs}), and describe how we curate the dataset of \javascript projects (Section~\ref{method:data}).
We present the analysis procedure used to extract reclassification events and construct a practice taxonomy (Section~\ref{method:pipeline}), which we use to address these questions in Sections~\ref{rq:rq1}--\ref{rq:rq4}.

\subsection{Research Questions}\label{method:rqs}
\edit{
We present the research questions using the Goal--Question--Metric (GQM) paradigm~\citep{basili1994gqm}.
The goal of this study is to characterize dependency reclassification with respect to prevalence, structural forms, and timing.
We study dependency role histories reconstructed from \package revisions in the context of \javascript projects that actively maintain their dependencies.
We address this goal through four research questions.
}

\edit{
\begin{description}
    \item[RQ1:] \RQone
    This question establishes how common reclassification is.
    We answer it with the share of projects that reclassify their dependencies at least once, and with project level and dependency level shares of each practice by starting role (Table~\ref{tab:practice_distribution}).
    \item[RQ2:] \RQtwo
    This question characterizes how removal contributes to the structural forms of reclassification.
    We answer it with removal size (batch versus singleton), introduction timing and family cohesion within batches, partial versus full reversion, and intermediate role paths before deletion.
    \item[RQ3:] \RQthree
    This question characterizes how role reassignment contributes to the structural forms of reclassification.
    We answer it with the forms of reassignment among dependency roles, covering single-direction transitions and role oscillation.
    \item[RQ4:] \RQfour
    This question establishes how long reclassification takes, completing the timing dimension of the goal.
    We answer it with how long each reclassification practice takes across roles, and with survival estimates of time to first reclassification and of return to the original role.
\end{description}
}

\begin{figure}[t]
    \centering
    \includegraphics[width=\linewidth]{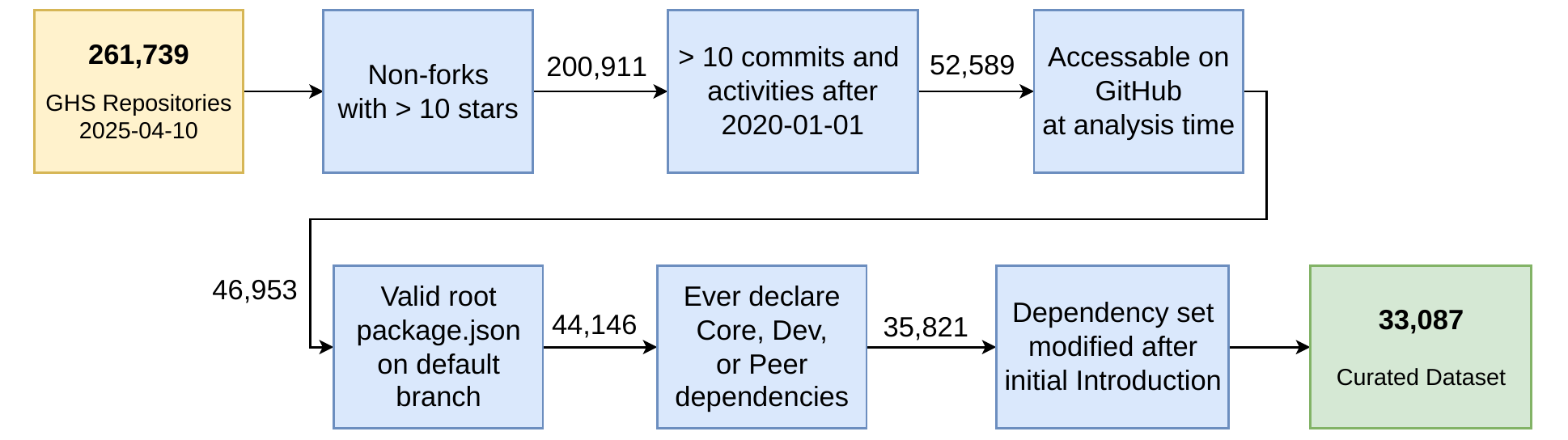}
    \caption{\small
    Project selection pipeline.
    Starting from the GHS dataset, each step reports the number of repositories retained after applying the corresponding filter.
    The final curated dataset of \totalprojects projects is the input to the analysis procedure in Section~\ref{method:pipeline}.
    }
    \label{fig:project_selection}
\end{figure}
\bigskip

\subsection{Project Selection}\label{method:data}
We start from 261{,}739 repositories retrieved via the GitHub Search Engine (GHS) dataset \citep{dabic2021sampling} (2025-04-10).
Following prior work \citep{soto2021longitudinal, he2023automating}, we exclude forks and include repositories with more than 10 stars, leaving 200{,}911 candidates.

To reduce noise from very small repositories and to focus on projects with a non-trivial development history, we require each repository to have more than 10 commits and at least one commit after \texttt{2020-01-01}, yielding \totaloriprojects candidates.
At the time of analysis (2025-04-16), \nolongerprojects of these repositories were no longer accessible, leaving 46{,}953 projects.

Because many repositories use multiple programming languages or do not use \package as a project manifest, we retain only repositories whose default branch contains a syntactically valid \package file in the latest commit.
We verify syntactic validity by parsing each retrieved snapshot with Python's standard \texttt{json} library.
A snapshot is accepted only if it parses as a well-formed JSON object; snapshots that fail to parse are excluded from analysis.
This filter excludes 2,807 projects and 44,146 projects remain.
\bigskip
\begin{table}[t]
    \centering
    \small
    \setlength{\tabcolsep}{15pt}
    \caption{\small
    Descriptive statistics of the curated dataset (\totalprojects projects).
    }
    \label{tab:dataset_stats}
    \begin{tabular}{lrrrrr}
        \toprule
        \textbf{Metric} & \textbf{Min} & \textbf{Median} & \textbf{Mean} & \textbf{Max} \\
        \midrule
        Stars                 & 11      & 46      & 521      & 320{,}950  \\
        Forks                 & 0       & 12      & 91       & 72{,}369  \\
        Commits               & 11      & 113     & 437      & 77{,}107 \\
        Project size (KB)    & 6       & 1{,}213 & 12{,}840 & 997{,}264 \\
        \bottomrule
    \end{tabular}
\end{table}

We exclude repositories that never declare any dependencies under \texttt{dependencies}, \texttt{devDependencies}, or \texttt{peerDependencies} throughout the history of the default branch.
To test this condition, we traverse the default-branch history of each candidate repository and check, at every commit that modifies a root \package file, whether at least one of the three sections contains at least one dependency name.
Repositories with only empty or missing dependency sections across the full history are excluded.
After this filter, 35{,}821 \javascript projects remain.

Furthermore, from these projects, we exclude 2{,}734 cases where the declared dependency set was never modified after its initial introduction. 
After this step, \totalprojects projects remain in our curated dataset.
\autoref{tab:dataset_stats} reports descriptive statistics of our dataset.
These statistics (mean exceeding median across all metrics) indicate that the corpus primarily captures routine dependency maintenance in both ordinary community projects, and highly popular and large-scale ecosystem projects.

% We identify such projects using the reclassification detection procedure described in Section~\ref{method:reclass}.
% Together, they account for 89\,\% of the 35{,}821 \javascript projects in our initial dataset, leaving only 11\,\% that exhibit no dependency reclassification over time.

%At the time of analysis (2025-04-16),  32.3\,\% (\recentprojects) of projects in our dataset have at least one commit after 2023-01-01, indicating a substantial number of recently maintained projects.

\subsection{Analysis Procedure}\label{method:pipeline}
\autoref{fig:method-overview} presents the analysis procedure applied on the curated dataset.
The curated dataset of \totalprojects projects serves as the input.
The procedure then proceeds in four steps: constructing dependency manifest histories (Section~\ref{method:history}), identifying reclassification events (Section~\ref{method:reclass}), aggregating events into reclassification sequences (Section~\ref{method:sequence}), and deriving the reclassification practice taxonomy (Section~\ref{method:practices}).
The resulting taxonomy forms the basis for the analyses that answer RQ1--RQ4 in Section~\ref{sec:results}.

We automate the procedure with Python scripts that retrieve \package snapshots from Git history, parse them with the standard \texttt{json} library, structurally compare consecutive snapshots using \texttt{jsondiff}, and aggregate the extracted events into sequences and practices.

\subsubsection{Constructing Dependency Manifest Histories}\label{method:history}
The first step constructs the dependency manifest history for each project.
We traverse the default branch and collect only the commits that modify a \package file; commits unrelated to the dependency manifest are excluded from the analysis.
For each such commit, we retrieve the corresponding \package snapshot, yielding an ordered sequence of manifest snapshots per project.
The resulting manifest histories serve as the input for identifying reclassification events.

\begin{figure}[t]
    \centering
    \includegraphics[width=\linewidth]{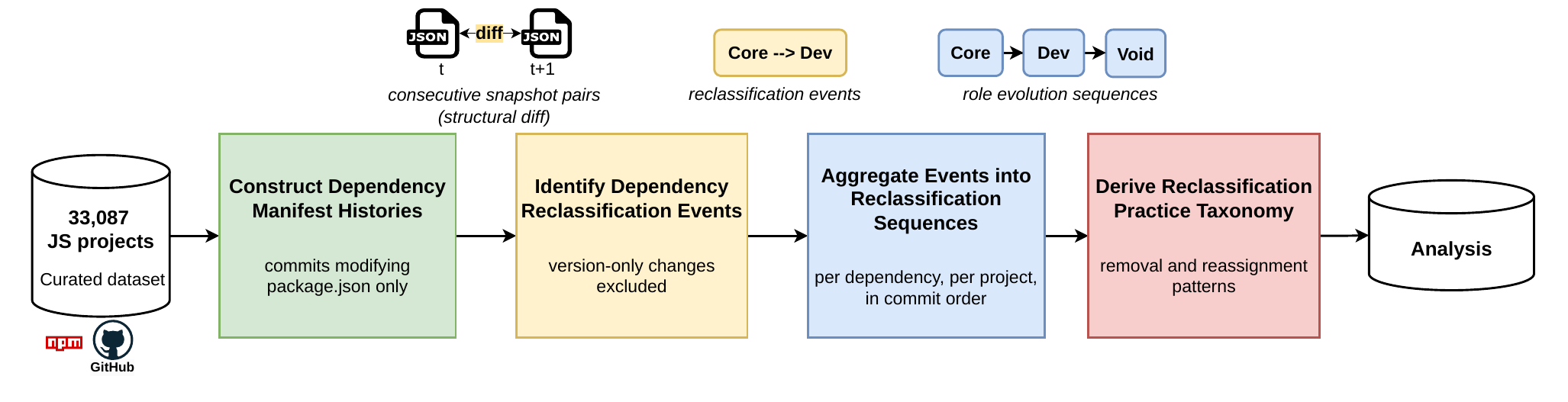}
    \caption{\small Workflow of the analysis procedure applied to the \totalprojects curated projects.}
    \label{fig:method-overview}
\end{figure}

\subsubsection{Identifying Dependency Reclassification Events}
\label{method:reclass}
We define a \textbf{reclassification event} as a change in the declared role of a dependency after its initial introduction into a project.
For each pair of consecutive \package snapshots in a project's manifest history (Section~\ref{method:history}), we compare the three dependency fields using structural differencing.
Each field denotes a declared role: \texttt{dependencies} for \core, \texttt{devDependencies} for \develop, and \texttt{peerDependencies} for \peer.
A dependency's role in a snapshot is determined by the field in which it is declared; if a field is absent from a snapshot, we treat it as containing no dependencies.

A deletion of a dependency name from one field together with its insertion into another yields a reclassification between the corresponding roles.
A deletion without insertion into another role yields a removal, whereas an insertion without deletion from another role represents an introduction or reintroduction.
We track dependencies by name, ignoring version strings:
a change that updates only a dependency version while leaving its declared role unchanged does not produce a role transition.

\begin{figure}[t]
    \centering
    \includegraphics[width=\linewidth]{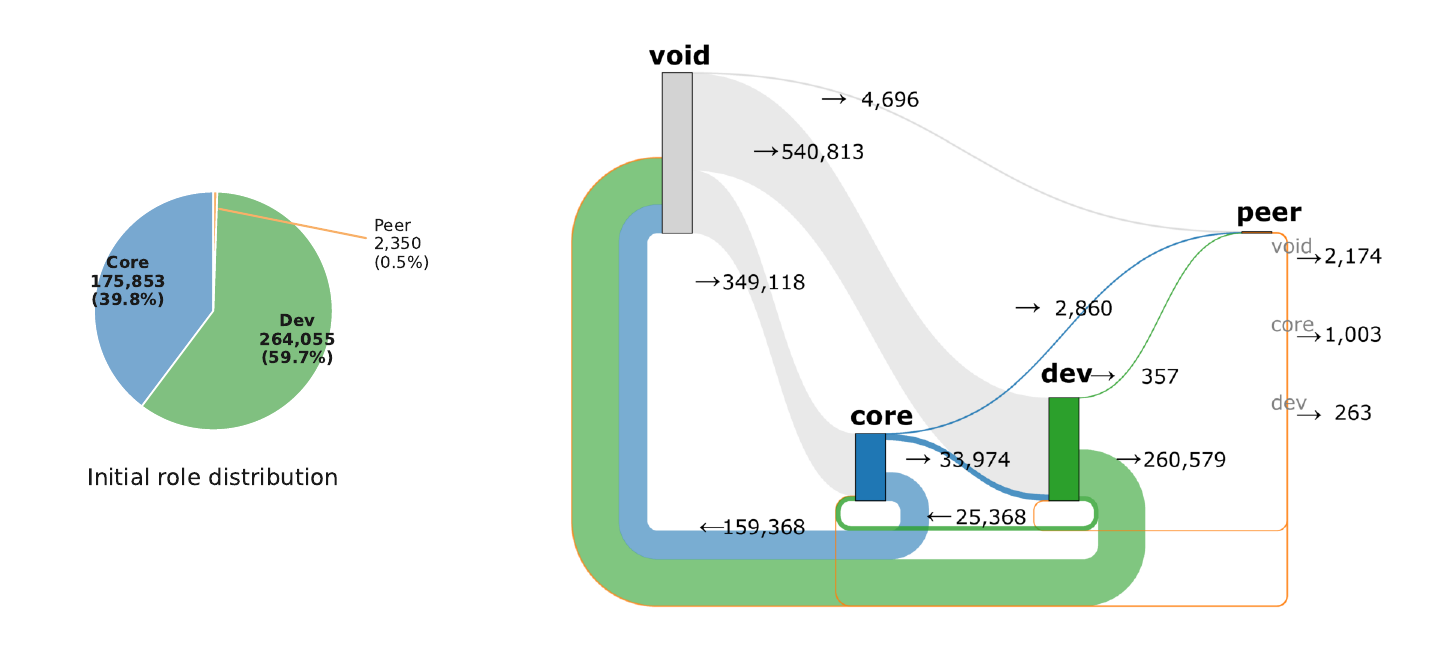}
    \caption{\small Left: initial distribution of dependencies by declared role (Core, Dev, Peer); Peer dependencies form only a small share of the initial set.
    Right: Sankey diagram of all dependency role transitions between consecutive \package snapshots.
    Each link represents a transition from a source role to a target role, with flow direction indicated by arrows and annotated with the absolute count.
    }
    \label{fig:sankey_reclassifications}
\end{figure}

To represent absence together with the three declared roles, we introduce an explicit \void state, which indicates that a dependency is not listed in any of the three fields.
This allows us to uniformly represent removals (e.g., \core \arrow \void) and reintroductions (e.g., \void \arrow \core) as role transitions.

Applying this differencing method across all consecutive snapshot pairs produces the complete set of observed role transitions.
\autoref{fig:sankey_reclassifications} summarizes their distribution across \totalprojects projects.
This includes 175,853 dependencies initially declared as Core (from 20,309 projects), 264,055 as Dev (from 20,328 projects), and 2,350 as Peer (from 1,276 projects).

Not every observed transition constitutes a reclassification event.
To isolate genuine reclassification events, we exclude two types of transitions:
(1) initial introductions, i.e., the first occurrence of a dependency name in the project history; and
(2) cross-file relocations in multi-package projects (e.g., monorepos), where a dependency is removed from one \package file and added to another within the same commit without changing its declared role.
After applying these exclusion rules, the remaining transitions form the set of reclassification events used in subsequent steps.

\autoref{fig:json_diff} illustrates the procedure on a concrete snapshot pair: depB is removed (\core \arrow \void), depC is reassigned (\core \arrow \develop), depD is a version-only update (excluded), and depE is an initial introduction (excluded).
Accordingly, we record two reclassification events for this diff.

\subsubsection{Aggregating Events into Reclassification Sequences}\label{method:sequence}
A single reclassification event captures one role transition (e.g., \core \arrow \develop\ or \core \arrow \void).
To examine how dependency declarations evolve over time, the third step aggregates events into reclassification sequences.

For each project, we group reclassification events by dependency name and order them according to their position in the default-branch history.
Each grouped case corresponds to one dependency within one project that undergoes at least one reclassification event after its initial introduction.
We construct a \textit{reclassification sequence} by taking the dependency's starting role and appending each later role reached through reclassification events, including reassignments, removals, and reintroductions represented through \void.
For example, a dependency introduced as \core, later reassigned to \develop, and eventually removed yields the sequence \core \arrow \develop \arrow \void.

Across all projects, we identify 442,258 dependency cases with at least one reclassification event after introduction.
These cases correspond to 533 distinct reclassification sequence patterns, which we group into broader practices in the next step.
\begin{figure}[t]
    \centering
    \includegraphics[width=0.7\linewidth]{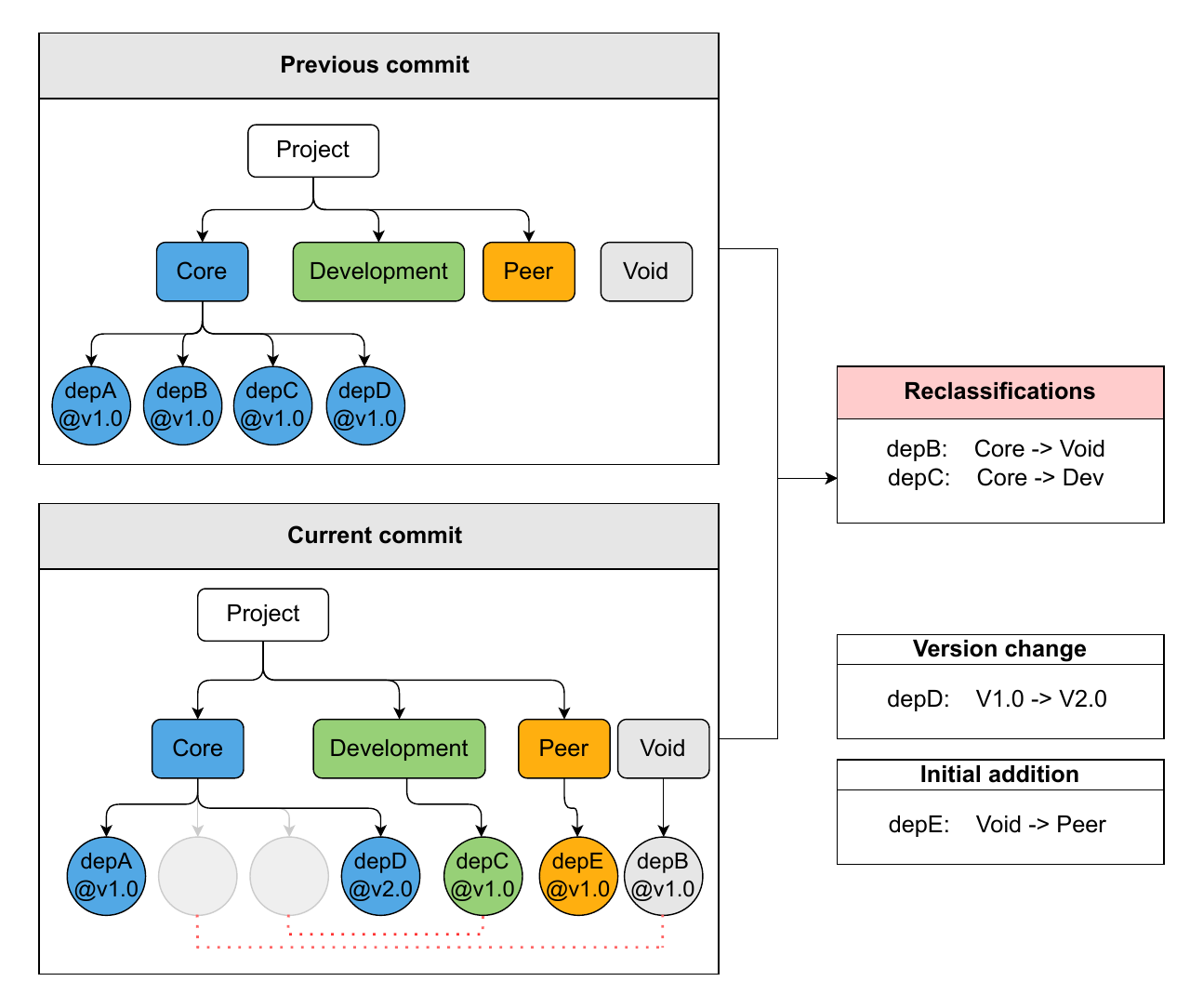}
    \caption{\small Identifying dependency reclassification events from two consecutive \package snapshots.}
    \label{fig:json_diff}
\end{figure}

\subsubsection{Deriving the Reclassification Practice Taxonomy}\label{method:practices}
The last step derives a taxonomy of reclassification practices from the 533 distinct sequence patterns.
We group sequences based on three observable properties:
(i) whether the sequence ends in absence (\void) or in a declared role,
(ii) whether intermediate role changes occur before the final state is reached, and
(iii) whether the dependency returns to its starting role after moving away from it.

Using these properties, we organize sequences into two broad categories: removal-related practices and reassignment-related practices.
\begin{itemize}
\item Removal practices.
These sequences are centered on deletion from the project.
They include \textit{one-step removal}, where the dependency moves directly from its starting role to \void\ (e.g., \core \arrow \void), and \textit{multi-step removal}, where deletion occurs only after one or more intermediate reassignments (e.g., \core \arrow \develop \arrow \void).
We also identify \textit{removal reversion}, where a dependency is removed and later reintroduced to its starting role without visiting another non-\void\ role in between (e.g., \develop \arrow \void \arrow \develop).

\item Reassignment practices.
These sequences retain the dependency in the project but revise its declared role.
They include \textit{role reassignment}, where the terminal role differs from the starting role (e.g., \core \arrow \develop\ or \peer \arrow \core), and \textit{role oscillation}, where the dependency moves away from its starting role and later returns to it after visiting at least one other non-\void\ role (e.g., \core \arrow \develop \arrow \core).

\end{itemize}

This taxonomy provides a structured view of how dependency declarations are revised after introduction.
It distinguishes practices by their structural characteristics (removal vs. reassignment, one-step vs. multi-step, reversion vs. oscillation) and forms the basis for analyzing the prevalence, structural forms, and temporal patterns of reclassification in Section~\ref{sec:results}.

\section{Results}\label{sec:results}
\subsection{\RQone}\label{rq:rq1}
We first examine the prevalence of dependency reclassification across \totalprojects dependency-active \javascript projects. 
Our results show that reclassification is a widespread practice: 79.1\,\% (26,187) of these projects undergo at least one dependency reclassification event, accounting for a total of \totalreccommits commits (19.4\,\% of all 
\totalcommits dependency-maintenance commits). 
In contrast, only 20.9\,\% (\totalonlyadd) of the projects limit their maintenance activities to version updates and new additions without ever reclassifying a dependency. 
Specifically, we observe reclassification across all role types: 20,309 projects revised 175,853 Core dependencies, 20,328 projects modified 264,055 Dev dependencies, and 1,276 projects adjusted 2,350 Peer dependencies.
These results show that dependency role declarations are often revised during project evolution rather than remaining fixed after introduction.

This result is also consistent with prior work \citep{zerouali2026comprehensive}, which reports continued dependency maintenance through additions and removals after long inactive periods. 
Our results extend this view by showing that such maintenance also frequently involves reclassification across declared roles during routine maintenance activities.
In the remainder of the results, we restrict the analysis to the 79.1\,\% of dependency-maintaining projects that exhibit at least one reclassification event, since the remaining 20.9\,\% provide no observable reclassification behavior.

Based on the taxonomy defined in Section~\ref{method:practices}, we distinguish removal-related and reassignment-related practices and analyze their prevalence at both the project and dependency levels.
\autoref{tab:practice_distribution} summarizes the distribution of these practices.

\begin{table*}[t]
    \centering
    \scriptsize 
    \caption{Distribution of reclassification practices across dependency roles. Columns are grouped by the overall reclassifying projects and the \textbf{initially declared role} (Core, Dev, Peer). 
    For each role, \textit{Projs} reports the percentage of projects in which the practice occurs at least once. 
    \textit{Deps} reports the percentage of dependencies (additive to 100\%). }
    \label{tab:practice_distribution}
    \setlength{\tabcolsep}{2pt}
    
    \begin{tabular}{l c c c c c c c}
        \toprule
        & \textbf{Overall} & \multicolumn{2}{c}{\textbf{Core}} & \multicolumn{2}{c}{\textbf{Dev}} & \multicolumn{2}{c}{\textbf{Peer}} \\
        \cmidrule(lr){2-2} \cmidrule(lr){3-4} \cmidrule(lr){5-6} \cmidrule(lr){7-8}
        \textbf{Practice} & \textbf{Projs} & \textbf{Projs} & \textbf{Deps} & \textbf{Projs} & \textbf{Deps} & \textbf{Projs} & \textbf{Deps} \\
        & \textit{(26,187)} & \textit{(20,309)} & \textit{(175,853)} & \textit{(20,328)} & \textit{(264,055)} & \textit{(1,276)} & \textit{(2,350)} \\
        \midrule

        \textit{Removal Practices} & \textit{ 97.2\%} & & & & & & \\
        One-step removal    & 96.0\% & 89.4\% & 74.2\% & 95.1\% & 84.0\% & 65.5\% & 57.3\% \\
        Removal reversion   & 33.1\% & 22.4\% & 7.2\%  & 28.6\% & 7.3\%  & 10.9\% & 7.6\%  \\
        Multi-step removal  & 17.8\% & 16.1\% & 5.3\%  & 10.6\% & 2.4\%  & 7.8\%  & 6.3\%  \\
        \addlinespace

        \textit{Reassignment Practices} & \textit{38.0\%} & & & & & & \\
        Reclassify to Core  & 14.0\% & --     & --     & 17.5\% & 3.8\%  & 18.9\% & 16.0\% \\
        Reclassify to Dev   & 23.1\% & 29.4\% & 9.8\%  & --     & --     & 12.1\% & 8.7\%  \\
        Reclassify to Peer  & 5.3\%  & 6.1\%  & 1.1\%  & 1.0\%  & 0.1\%  & --     & --     \\
        Role Oscillation    & 11.2\% & 7.5\%  & 2.4\%  & 9.1\%  & 2.3\%  & 5.2\%  & 4.2\%  \\
        \midrule
        \textbf{Total}      & --     & --     & \textbf{100.0\%} & -- & \textbf{100.0\%} & -- & \textbf{100.0\%} \\
        \bottomrule
    \end{tabular}
\end{table*}

Across all roles, removal is prevalent (97.2\,\%) across projects.
Particularly, one-step removal appears in 89.4\,\% of projects with Core dependencies and 95.1\,\% of projects with Dev dependencies; at the dependency level, it accounts for 74.2\,\% of Core cases and 84.0\,\% of Dev cases. 
Although Peer dependencies exhibit lower prevalence, this practice still affects 65.5\,\% of projects with Peer dependencies and 57.3\,\% of Peer cases. 
Our result shows that direct removal is a widespread maintenance action across dependency roles.

Removal is not always final.
Removal reversion occurs in 22.4\,\% of projects with Core dependencies and 28.6\,\% of projects with Dev dependencies, with a lower but still notable share of 10.9\,\% for Peer dependencies.
This means that previously removed dependencies are often restored in later commits.
For example, in \texttt{scriptpilot/app-framework},\footnote{\href{https://github.com/scriptpilot/app-framework}{https://github.com/scriptpilot/app-framework}} a large batch of dependencies was removed in commit \href{https://github.com/scriptpilot/app-framework/commit/eb03d370}{\texttt{eb03d370}} (2018-12-29, message: \texttt{clean up}), where 61 Core dependencies and three Dev dependencies were deleted from the \package file. 
The same commit also bumped the package \texttt{version} field to \texttt{3.0.0}, so the cleanup coincided with a major version change.
Two days later, commit \href{https://github.com/scriptpilot/app-framework/commit/6e1b186d}{\texttt{6e1b186d}} (2018-12-31, message: \texttt{Revert ``clean up''}) explicitly reverted this change, restoring all previously removed dependencies. 
This sequence illustrates a removal reversion pattern in which a large dependency cleanup is later fully reversed in a subsequent revision.

Removal does not always occur as a single direct change.
Multi-step removal appears in 16.1\,\% of projects with Core dependencies, 10.6\,\% with Dev dependencies, and 7.8\,\% with Peer dependencies, indicating that deletion may be preceded by an intermediate role change.

Beyond removal, reassignment is also common (38\,\%) across projects.
At the project level, Core-to-Dev reclassification appears in 29.4\,\% of projects, Dev-to-Core in 17.5\,\% of projects, and Peer-to-Core in 18.9\,\% of projects. 
At the dependency level, Peer \arrow Core and Peer \arrow Dev account for 16.0\,\% and 8.7\,\% of Peer dependencies, respectively, while Core \arrow Dev represents 9.8\,\% of Core dependencies.
These results show that dependencies are not only deleted, but are also frequently retained in the project under revised declared roles.

Some of these role changes do not stabilize after a single reassignment, with 11.2\,\% of projects experiencing role oscillation.
It appears in 7.5\,\% of projects with Core dependencies, 9.1\,\% with Dev dependencies, and 5.2\,\% with Peer dependencies.
Dependencies move away from their original roles and later return, indicating that role adjustment may remain unsettled over time.

\edit{
Taken together, the dependency level shares in \autoref{tab:practice_distribution} differ across roles: removal remains dominant for all three, but Peer dependencies are more often reassigned to Core or Dev, and Core and Dev differ most on one-step removal and Core-to-Dev reassignment.
To statistically validate these differences, we apply Pearson's chi-squared test\footnote{The scripts used to compute all statistical tests and effect sizes reported in this paper are available at \href{https://github.com/ASSERT-KTH/dependency-reclassification/blob/main/scripts/analysis/compute_inferential_stats.py}{https://github.com/ASSERT-KTH/dependency-reclassification/compute\_inferential\_stats.py}.} to assess whether practice category is independent of dependency role, following prior work~\citep{chinthanet2021lags}.
To measure the strength of that association, we use Cram{\'{e}}r's $V$.
Following Cohen~\citep{cohen1988statistical}, for pairwise contrasts between two roles ($df^{*}{=}1$), we interpret $V{<}0.10$ as negligible, $0.10{\leq}V{<}0.30$ as small, $0.30{\leq}V{<}0.50$ as medium, and $V{\geq}0.50$ as large.
Practice category is associated with dependency role ($p{<}0.001$, $V{=}0.12$).
The strongest pairwise association is between Core and Dev ($V{=}0.16$, small), followed by Dev and Peer ($V{=}0.10$, small) and Core and Peer ($V{=}0.05$, negligible); all pairwise $p{<}0.001$.
Consistent with the role differences reported above, the largest contrast is between Core and Dev.
}
\bigskip

\begin{answerbox}
\textbf{Answer to RQ1:}
Across our dataset, 79.1\,\% of projects reclassify dependencies at least once.
Removal is the most prevalent practice (97.2\,\%), but it is not always final or direct: 33.1\,\% later reintroduce removed dependencies, and 17.8\,\% reassign dependencies before removal.
Beyond removal, reassignment across roles occurs in 38.0\,\% of projects, and 11.2\,\% of projects further exhibit role oscillation.
These findings confirm that dependency roles are actively maintained and reclassified through diverse forms throughout project evolution.
\end{answerbox}

\subsection{\RQtwo}\label{rq:rq2}
RQ1 shows that removal is the most prevalent form of dependency reclassification.
However, dependency deletion occurs in multiple distinct patterns.
Some dependencies are deleted directly and never reintroduced (we term this \textit{one-step removal}), some are later reintroduced after deletion (removal reversion), and others are removed only after an earlier role reassignment to another role (multi-step removal).
To answer this research question, we analyze these removal-related practices to understand how dependency deletion is carried out in practice.

\subsubsection{One-step removal}\label{prac:one_step}
One-step removal captures cases where a dependency is deleted directly from its declared role and is not reintroduced later in the observed project history.
It is the most common removal practice, occurring in 89.4\,\% of projects with Core dependencies, 95.1\,\% of projects with Dev dependencies, and 65.5\,\% of projects with Peer dependencies.
To better understand how one-step removals are performed in practice, we examine removal size, measured as the number of dependencies deleted together in the same commit, as shown in \autoref{fig:onetime_removal}.

We observe clear differences across dependency roles: Core and Dev dependencies are more often removed in batches, whereas Peer dependencies are more often removed individually.
In particular, 61.9\,\% of one-step Core removals and 79.6\,\% of one-step Dev removals occur in commits that delete at least two dependencies (size $\ge 2$).
Notably, for Dev dependencies, more than half of one-step removals (54.9\,\%) occur in large batches (size $\ge 5$).
By contrast, Peer removals are predominantly isolated, with 60.0\,\% affecting a single dependency.
This contrast suggests that one-step removal often serves as a broad cleanup action for Core and Dev dependencies, whereas Peer removal more often reflects targeted adjustment of individual dependencies.

\begin{figure}
    \centering
    \includegraphics[width=\linewidth]{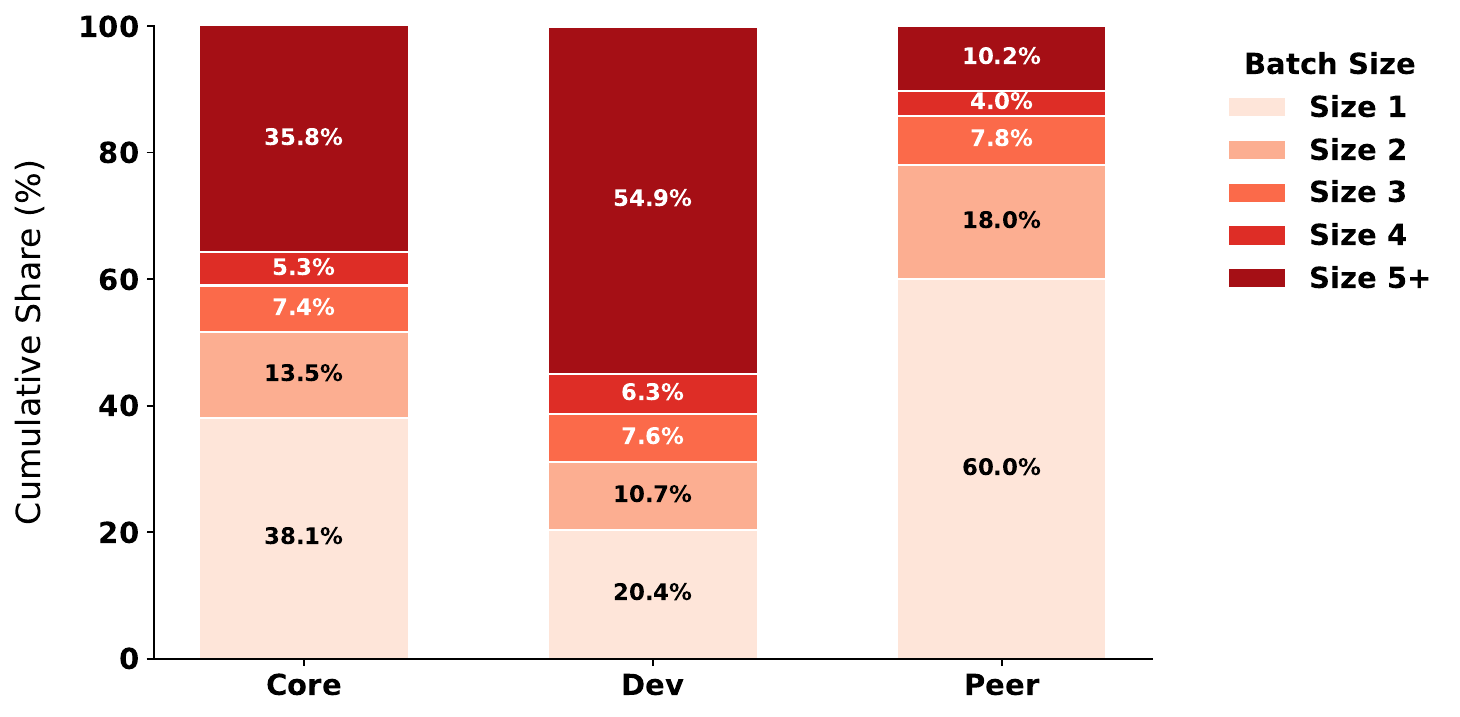}
    \caption{Distribution of one-step removal by the number of dependencies deleted in the same commit, shown separately for Core, Dev, and Peer dependencies.
    }
    \label{fig:onetime_removal}
\end{figure}

These size differences raise a further question: when developers delete many dependencies in one commit, what maintenance activity are they performing?
Given the prevalence of batch removals, we examine whether dependencies deleted in the same commit share a single introduction commit.

Across Core and Dev dependencies, most removal batches (size $\ge 2$) combine dependencies adopted at different points in project history (61.0\,\% for Core and 58.8\,\% for Dev).
The remaining batches share a single introduction commit: they delete a batch introduced in a single earlier commit.
For example, after dependencies were copied from a template, maintainers may later delete that whole set once they recognize it is unnecessary for the package.
In \texttt{ember-cli/ember-export-application-global},\footnote{\href{https://github.com/ember-cli/ember-export-application-global}{https://github.com/ember-cli/ember-export-application-global}} commit~\href{https://github.com/ember-cli/ember-export-application-global/commit/7e3b787}{\texttt{7e3b787}} (2014-10-09, message: \texttt{Update package.json}) added 11 Core dependencies in one change.
About two hours later, commit~\href{https://github.com/ember-cli/ember-export-application-global/commit/d2b96b4}{\texttt{d2b96b4}} (message: \texttt{Fix stupid dependency copying.}) removed all 11.
The commit message records that these Core declarations had been copied and were not required by the project itself.

Batches that remove packages introduced in different commits are more common.
They often correspond to an accumulated cleanup: packages added across earlier stages of the project are deleted together in one later commit.
For example, \texttt{cerner/terra-ui}\footnote{\href{https://github.com/cerner/terra-ui}{https://github.com/cerner/terra-ui}} is a documentation site for a UI component library.
In pull request~\href{https://github.com/cerner/terra-ui/pull/600}{\#600} (commit~\href{https://github.com/cerner/terra-ui/commit/ba56fdd}{\texttt{ba56fdd}}, 2023-11-13, message: \texttt{Package json cleanup}), maintainers removed 72 Core dependencies that had been introduced across 13 earlier commits spanning years of project history.
The pull request explains that many of these packages were not used directly in the repository: some were already installed automatically as peer dependencies, and others were redundant because they were already pulled in as transitive dependencies of the documentation packages.
This case shows an accumulated Core cleanup in which long-lived declarations from different introduction commits are removed once maintainers decide they are no longer needed as direct dependencies.

Introduction timing alone does not fully characterize a batch removal.
We therefore also examine whether the deleted packages form a cohesive dependency family.
For batches of size $\ge 3$, we identify family-dominated removals in which at least 50\,\% of removed dependencies share a common namespace or prefix (e.g., \texttt{eslint-} or \texttt{@babel}).
Family-dominated batches appear more frequently in Dev removals (58.1\,\%) than in Core removals (35.2\,\%).
For Dev, this second perspective points to toolchain retirement: related build or development packages are deleted together when that toolchain is abandoned.
For example, \texttt{fengyuanchen/cropper}\footnote{\href{https://github.com/fengyuanchen/cropper}{https://github.com/fengyuanchen/cropper}} is an image-cropping \javascript library.
In commit~\href{https://github.com/fengyuanchen/cropper/commit/d7278ca}{\texttt{d7278ca}} (2015-08-09, message: \texttt{Remove Grunt tasks}), maintainers deleted the project's \texttt{Gruntfile.js} and removed 19 Dev dependencies, all from the Grunt toolchain (e.g., \texttt{grunt}, \texttt{grunt-contrib-uglify}, and \texttt{grunt-contrib-watch}).
No replacement Dev packages were added in the same commit.
This case shows a family-dominated Dev batch used to retire an entire development toolchain in one change.

Taken together, these two perspectives separate different batch-removal activities.
Along introduction timing, some batches remove an earlier copied batch, while most Core batches remove packages that were introduced in different commits and thus match accumulated cleanup.
Along family cohesion, Dev batches are more often family-dominated, and the cases we examined show toolchain retirement, where related development packages are removed together with the workflow that required them.

\edit{
We apply Pearson's chi-squared test to assess whether batch versus singleton removal is independent of dependency role.
The strongest pairwise association is between Core and Dev ($V{=}0.20$, small), followed by Dev and Peer ($V{=}0.08$, negligible) and Core and Peer ($V{=}0.05$, negligible); all pairwise $p{<}0.001$.
Pearson's chi-squared test for family-dominated versus other batches likewise shows an association between Core and Dev ($V{=}0.21$, small, $p{<}0.001$).
Consistent with the differences reported above, Core and Dev differ most clearly in how often one-step removal is batched, and Dev batches are more often family-dominated.
}

\subsubsection{Removal reversion}
\label{pattern:rq2_removal_reversion}
Although one-step removal may delete batches introduced earlier, it does not involve later reintroduction.
Removal reversion studies cases where dependencies are removed and later reintroduced to the same role.
It is observed across all roles (Table~\ref{tab:practice_distribution}), affecting 22.4\,\% of projects with Core dependencies, 28.6\,\% of projects with Dev dependencies, and 10.9\,\% of projects with Peer dependencies.

To understand how removals are later revised, we examine whether reintroduction restores the full earlier removal or only part of it.
In particular, we identify 1,133 reversions after batch removals (size $\ge 2$), and in 62\,\% of these reversions, only a subset of the dependencies is reintroduced.
This share increases for larger batches: in 558 reversions after batch removals with size $\ge 3$, it reaches 71.3\,\%.
Correspondingly, the remaining 38\,\% of batch reversions restore the entire batch (full removal reversion).

As an illustration of full removal reversion, consider \texttt{folio-org/stripes-core},\footnote{\href{https://github.com/folio-org/stripes-core}{https://github.com/folio-org/stripes-core}} a shared UI package for an open-source library management system.
In pull request~\href{https://github.com/folio-org/stripes-core/pull/1003}{\#1003} (commit~\href{https://github.com/folio-org/stripes-core/commit/6d2d1f1}{\texttt{6d2d1f1}}, 2021-02-01, message: \texttt{STRIPES-730: remove webpack from stripes-core}), maintainers moved the Webpack setup into a separate package and deleted 62 Core dependencies related to that build toolchain.
The next day, pull request~\href{https://github.com/folio-org/stripes-core/pull/1010}{\#1010} (commit~\href{https://github.com/folio-org/stripes-core/commit/5783e75}{\texttt{5783e75}}, 2021-02-02, message: \texttt{Revert ``STRIPES-730: remove webpack from stripes-core''}) restored all 62 dependencies.
The revert discussion reports compatibility problems with \texttt{stripes-cli} and related packages, so the maintainers rolled the entire batch back.

Some partial removal reversions are intentional adaptations that are revisited as the project evolves.
In \texttt{cerner/terra-dev-site},\footnote{\href{https://github.com/cerner/terra-dev-site}{https://github.com/cerner/terra-dev-site}} a library in a UI framework that builds a documentation website for its components, pull request~\href{https://github.com/cerner/terra-dev-site/pull/70}{\#70} (commit~\href{https://github.com/cerner/terra-dev-site/commit/402c152}{\texttt{402c152}}, 2018-04-23, message: \texttt{Consume Terra-Toolkit Webpack~4}) migrated the Webpack pipeline to the shared \texttt{terra-toolkit} configuration and removed 11 Core dependencies, including several Webpack loaders.
Nearly three years later, after \texttt{terra-toolkit} was removed from the project's peer dependencies,\footnote{\href{https://github.com/cerner/terra-dev-site/pull/337}{PR~\#337}; changelog for v6.34.0: \textit{Removed terra-toolkit from peer dependency}.} those loaders were again required.
In the same-day hotfix \href{https://github.com/cerner/terra-dev-site/pull/340}{PR~\#340} (commit~\href{https://github.com/cerner/terra-dev-site/commit/8a3434a}{\texttt{8a3434a}}, 2021-03-26, message: \texttt{Fix loaders problem}), only five of the earlier removed dependencies were restored to Core; the changelog states that the loaders were ``added back \ldots{} due to the terra-toolkit removal.''
That commit also updated the package \texttt{version} field (to \texttt{6.34.1}), so the partial reintroduction was shipped as a patch release.

Overall, removal reversion represents an iterative correction process in dependency maintenance: a removal is not always a durable end state, but a decision that may be revised.
The prevalence of partial reversion reflects both imperfect cleanup and ongoing adaptation of dependency declarations over time.

\subsubsection{Multi-step removal}
\label{prac:multi_step}
Unlike one-step removal, where deletion occurs directly from the starting role, 
multi-step removal involves intermediate reassignment.
Specifically, it captures cases where a dependency is reassigned to at least 
one intermediate role before its eventual removal.

Across roles, multi-step removals are most often characterized by a single intermediate reassignment before deletion.
In 82.8\,\% of Core cases, dependencies are reassigned to Dev before being removed. 
Similarly, in 85.4\,\% of Dev cases, dependencies are reassigned to Core before being removed. 
Peer dependencies follow more varied routes, primarily passing through either Core (45.3\,\%) or Dev (29.7\,\%) before their final removal.

These path statistics raise a further question: why do developers reassign a dependency before deleting it?
The following cases show that the intermediate reassignment and the later deletion are often separate maintenance needs.

One common scenario is Core-to-Dev reassignment followed by later deletion.
For example, \texttt{code-dot-org/craft}\footnote{\href{https://github.com/code-dot-org/craft}{https://github.com/code-dot-org/craft}} is a game engine used in an educational coding platform.
In pull request~\href{https://github.com/code-dot-org/craft/pull/40}{\#40} (commit~\href{https://github.com/code-dot-org/craft/pull/40/commits/ee7dbdfa671ec0e99d56370bf6b49b9bcf753066}{\texttt{ee7dbdf}}, 2017-08-17, message: \texttt{Add Travis CI}), maintainers moved build-related packages such as \texttt{babelify} and \texttt{browserify} from Core to Dev while setting up CI.
This step revised installation scope: build-time plugins that had remained in Core for nearly two years (712 days) were placed under Dev, while the project continued to use Browserify as its bundler.
Those packages then remained in Dev for another 309 days.
They were removed only in pull request~\href{https://github.com/code-dot-org/craft/pull/467}{\#467} (commit~\href{https://github.com/code-dot-org/craft/pull/467/commits/7cd3130}{\texttt{7cd3130}}, 2018-06-22, message: \texttt{Build with webpack}), when the project replaced Browserify with Webpack.
In this case, each step is a different maintenance need: \core \arrow \develop\ fixes production install scope under the Browserify toolchain, and \develop \arrow \void\ later retires those packages only when the toolchain itself is replaced.

The opposite path also appears: Dev-to-Core reassignment followed by later deletion.
In \texttt{automattic/wp-calypso},\footnote{\href{https://github.com/Automattic/wp-calypso}{https://github.com/Automattic/wp-calypso}} a web application for managing WordPress.com sites, maintainers migrated the repository from \npm to Yarn (another \javascript package manager) and enabled Yarn workspaces.
As part of that migration, pull request~\href{https://github.com/Automattic/wp-calypso/pull/40880}{\#40880} (commit~\href{https://github.com/Automattic/wp-calypso/commit/c9a8c34}{\texttt{c9a8c34}}, 2020-04-08, message: \texttt{Moves existing devDependencies to dependencies}) moved 146 packages from Dev to Core.
The migration plan~\href{https://github.com/Automattic/wp-calypso/issues/40882}{\#40882} explains why: with packages left in Dev, Yarn workspaces could fail during install (a known Yarn bug~\href{https://github.com/yarnpkg/yarn/issues/7807}{yarnpkg/yarn\#7807}).
Moving the packages to Core was a workaround to make Yarn workspaces install succeed despite the known bug; the PR does not claim that production runtime actually required them.
About seven months later, pull request~\href{https://github.com/Automattic/wp-calypso/pull/46515}{\#46515} (commit~\href{https://github.com/Automattic/wp-calypso/commit/4d80c49}{\texttt{4d80c49}}, 2020-10-23, message: \texttt{Delete unused dependencies}) deleted many of them from the root \texttt{package.json} as unused there.
Because the constraint was specific to Yarn workspaces during migration, maintainers temporarily broadened the production dependency set to get the install working, and then cleaned up those packages once the workspace layout stabilized.

Overall, multi-step removal reveals that dependencies can be deleted from a different context than where they were originally declared.
The intermediate reassignment and the eventual removal are often driven by different needs, which is why reassignment should also be analyzed as a distinct maintenance practice in the following section.
\bigskip

\begin{answerbox}
\textbf{Answer to RQ2:}
Developers use distinct removal practices:
One-step removal often appears as batch cleanup for Core and Dev dependencies (61.9\,\% of Core and 79.6\,\% of Dev one-step removals delete $\ge 2$ dependencies), whereas 60.0\,\% of Peer dependencies are removed individually.
When developers reverse a previous batch removal, they reintroduce only a subset of the removed dependencies in 62\,\% of cases.
About 80\,\% of multi-step removals involve a single intermediate role reassignment before final deletion, confirming that the dependency’s declared role can change prior to its removal.
Together, these practices show removal as staged maintenance: batch cleanup, role change before deletion, and only partial restore when a prior batch proves incomplete.
\end{answerbox}

\subsection{\RQthree}\label{rq:rq3}
RQ3 examines role reassignment practices, that is, cases where developers reclassify dependencies across declared roles without removing them.
Unlike removal practices discussed in RQ2, these reclassifications preserve the dependency in the project while revising the role in which it is installed and maintained.
This question therefore focuses on how developers reorganize dependency boundaries across runtime, development, and consumer-managed contexts.
As summarized in Table~\ref{tab:practice_distribution}, reassignment is common across all original roles, with 29.4\,\% of projects engaging in Core-to-Dev reclassification, 17.5\,\% performing Dev-to-Core reclassification, and 18.9\,\% practicing Peer-to-Core reclassification.

\subsubsection{Reclassification between Core and Dev}\label{prac:core_dev}
Reclassification between Core and Dev is the most prominent form of role reassignment.
Core-to-Dev reclassification affects 9.8\,\% of Core dependencies and appears in nearly 30\,\% of projects.
This reclassification removes the dependency from production installs while retaining it for development, thereby narrowing the project’s production install surface and reducing runtime dependency bloat.

To characterize the dependencies most frequently reclassified from Core to Dev, we first examine how concentrated these reclassifications are.
We find that Core-to-Dev reclassification is highly concentrated among a relatively small set of dependencies.
Specifically, 75\,\% of all such reclassifications are contributed by 176 dependencies, affecting 81.1\,\% of involved projects.
To understand the functional roles of these key dependencies, we manually inspected their documentation and grouped them by primary functionality.
This yields three broad categories: development infrastructure, general utilities, and runtime framework libraries, summarized in \autoref{tab:core_to_dev_categories}.

\begin{table}[t]
\centering
\scriptsize
\caption{Category breakdown of dependencies undergoing Core-to-Dev reclassifications: top 176 dependency families (covering 75.0\% of all cases)}
\label{tab:core_to_dev_categories}
\begin{tabular}{llrrl}
\toprule
\textbf{Major Group} & \textbf{Category} & \textbf{Cases} & \textbf{\%} & \textbf{Examples} \\
\midrule
\multirow{5}{*}{Infrastructure} 
    & Transpilers \& Compilers    & 5,399 & 16.2 & babel, postcss, sass \\
    & Build \& Task Tooling       & 3,606 & 10.9 & webpack, gulp, rollup \\
    & Linting \& Formatting       & 2,575 & 7.7  & eslint, prettier \\
    & Testing Frameworks          & 1,933 & 5.8  & jest, mocha, chai \\
    & Type Declarations           & 1,049 & 3.2  & @types \\
\midrule
Utility 
    & General-purpose Utility     & 5,302 & 16.0 & lodash, chalk \\
\midrule
\multirow{2}{*}{Runtime} 
    & UI Frameworks \& Ecosystem  & 4,525 & 13.6 & react, vue, angular \\
    & Backend \& Networking       & 534   & 1.6  & express, axios, request \\
\bottomrule
\end{tabular}
\end{table}
The largest group consists of development infrastructure, such as transpilers, bundlers, linters, and test frameworks (e.g., \texttt{babel}, \texttt{webpack}, and \texttt{eslint}).
These packages typically belong in Dev because they support building, testing, or analyzing the project rather than executing the application in production.
For example, \texttt{sensu/web}\footnote{\href{https://github.com/sensu/web}{https://github.com/sensu/web}} is a web UI for a monitoring platform.
In pull request~\href{https://github.com/sensu/web/pull/388}{\#388} (commit~\href{https://github.com/sensu/web/commit/648ee2f}{\texttt{648ee2f}}, 2021-11-09), maintainers moved 81 development-infrastructure dependencies from Core to Dev, including 15 transpilers and 25 build tools.
The commit message states the goal explicitly: \texttt{differentiate between dependencies used by the web app and those used in development}.
This case shows Core-to-Dev reclassification used to shrink the production dependency set and separate runtime packages from development tooling.
This is consistent with prior work showing that many dependencies declared as runtime are not actually used in production \citep{latendresse2022not}.

The second group consists of general-purpose utility libraries, such as \texttt{lodash}, \texttt{chalk}, and \texttt{commander}.
These libraries are not confined to one declared role and may move between Core and Dev over time.
For example, in \texttt{dhis2/d2-ui},\footnote{\href{https://github.com/dhis2/d2-ui}{https://github.com/dhis2/d2-ui}} a UI component library, \texttt{lodash} first entered as Core (\href{https://github.com/dhis2/d2-ui/commit/2cf86b27}{\texttt{2cf86b27}}, 2016-05-24, message: \texttt{Headerbar wip}).
It left Core when maintainers emptied the whole production dependency set (\href{https://github.com/dhis2/d2-ui/commit/c09afa39}{\texttt{c09afa39}}, 2016-11-14, message: \texttt{Update dependencies}).
It returned under Dev in the testing suite migration (\href{https://github.com/dhis2/d2-ui/pull/120}{PR~\#120}, commit~\href{https://github.com/dhis2/d2-ui/commit/1956778}{\texttt{1956778}}, 2017-09-20), where tests import \texttt{lodash}.
This case illustrates that as the project evolves, the need for the same dependency can change, and that change sometimes means a different scope.

The third group involves runtime framework libraries, such as \texttt{react} and \texttt{express}.
These cases often arise in projects that provide reusable components or plugins for those frameworks.
In such projects, the framework dependency may remain relevant for development and testing, while no longer being retained in the package’s own production declarations.

Conversely, Dev-to-Core reclassification expands the production install set: dependencies previously retained only for development are moved into the project's runtime dependency set.
This practice affects 3.8\,\% of Dev dependencies and appears in 17.5\,\% of projects.
Compared with Core-to-Dev reclassification (9.8\,\% of Core dependencies, nearly 30\,\% of projects), Dev-to-Core reclassification is less prevalent yet still non-negligible.

Interestingly, the dependency families most frequently reclassified in this direction substantially overlap with those in the Core-to-Dev direction, including \texttt{babel}- and \texttt{webpack}-related dependencies.
A representative example is \texttt{go-faast/faast-web},\footnote{\href{https://github.com/go-faast/faast-web}{https://github.com/go-faast/faast-web}} a web application for managing cryptocurrency portfolios.
In commit~\href{https://github.com/go-faast/faast-web/commit/48d8b42}{\texttt{48d8b42}} (2019-01-18, message: \texttt{Move everything out of devDependencies for production builds}), maintainers moved 119 dependencies from Dev to Core, including 11 \texttt{babel}-related and 25 \texttt{webpack}-related packages.
Only a small development-oriented set (e.g., \texttt{eslint} and \texttt{webpack-dev-server}) remained in Dev.
The change suggests that the application was built in a production environment where Dev dependencies were not installed. 
Maintainers moved nearly the entire Dev set into Core so that deployment would succeed, indicating a deployment requirement for the build toolchain itself.
The result is a production install that includes build tools such as Babel and Webpack alongside the actual runtime code, increasing install size and surface without a corresponding runtime need.
From a maintenance perspective, this is a temporary measure: deployment succeeded most reliably once build tooling was treated as production-installed code, even though that blurs the intended Dev/Core separation.

Taken together, the boundary between runtime and development is revised in both directions.
The same families of build tools may move from Core to Dev to reduce the production install set, or from Dev to Core when more packages must be installed, either for deployment needs (\texttt{faast-web}) or to satisfy package manager constraints during migration (\texttt{wp-calypso}).
These cases show ongoing adjustment of dependency installation scope.

\edit{We apply Pearson's chi-squared test to assess whether being reassigned to the other role is independent of the starting role (Core or Dev).
Core dependencies are more often reassigned to Dev than Dev dependencies are reassigned to Core ($V{=}0.12$, small effect; $p{<}0.001$), consistent with the shares reported above.}

\subsubsection{Reclassification of Peer Dependencies}\label{prac:peer}
Peer dependencies exhibit a higher rate of role reassignment compared to Core and Dev dependencies\edit{, as also reflected in the pairwise effect sizes reported in Section~\ref{rq:rq1}}.
As shown in \autoref{tab:practice_distribution}, 16.0\,\% of Peer reclassifications shift to Core, and 8.7\,\% to Dev, together accounting for nearly 25\,\% of all Peer reclassifications.
Since Peer dependencies are declared as requirements to be provided by consumers, these reclassifications shift them away from consumer-provided installation and into the project’s own dependency declarations.

One direction is Peer-to-Core reclassification.
In this direction, dependencies that were previously expected to be provided by consumers become part of the project’s own production dependency set.
For example, \texttt{wesbos/eslint-config-wesbos}\footnote{\href{https://github.com/wesbos/eslint-config-wesbos}{https://github.com/wesbos/eslint-config-wesbos}} is a shareable ESLint and Prettier configuration package.
Thirteen packages were initially declared as Peer dependencies (e.g., \href{https://github.com/wesbos/eslint-config-wesbos/commit/a0f49774559}{\texttt{a0f49774}}, 2019-02-26).
In pull request~\href{https://github.com/wesbos/eslint-config-wesbos/pull/125}{\#125} (commit~\href{https://github.com/wesbos/eslint-config-wesbos/commit/44001c0}{\texttt{44001c0}}, 2023-01-05, message: \texttt{Simplify install step}), they were reclassified to Core.
The PR discussion explains that Peer installation was hard for consumers and could force unnecessary installs.
The same commit also set the package \texttt{version} to \texttt{3.1.5-beta.4}, so the Peer-to-Core move was packaged with a prerelease bump.
This case shows Peer-to-Core reclassification used to move installation responsibility from consumers to the package itself and simplify adoption.
When consumers must install such dependencies manually, installation becomes more error-prone, consistent with the findings of \citep{wang2025understanding}.

The other direction is Peer-to-Dev reclassification.
In this direction, the dependency is no longer expected to be installed by consumers and is retained only for the project’s development workflow.
This is illustrated by \texttt{connor-baer/rich-text-to-jsx},\footnote{\href{https://github.com/connor-baer/rich-text-to-jsx}{https://github.com/connor-baer/rich-text-to-jsx}} a library that renders Contentful rich text as JSX.\footnote{JSX is an HTML-like syntax extension for JavaScript, commonly used with React to describe UI; see \url{https://react.dev/learn/writing-markup-with-jsx}.}
Initially, \texttt{react}, \texttt{react-dom}, and \texttt{prop-types} were all Peer dependencies (\href{https://github.com/connor-baer/rich-text-to-jsx/commit/7f61c10}{\texttt{7f61c10}}, 2018-12-26, message: \texttt{Initial implementation}).
In a testing refactor (\href{https://github.com/connor-baer/rich-text-to-jsx/commit/f7caf43}{\texttt{f7caf43}}, 2019-01-20, message: \texttt{Improve tests}), maintainers kept only \texttt{react} as Peer and moved \texttt{react-dom} and \texttt{prop-types} to Dev.
The split follows usage: consumers still need \texttt{react} to render the library's JSX, while \texttt{prop-types} is loaded only outside production and \texttt{react-dom} is brought in with test tooling such as \texttt{react-test-renderer}.
The maintainers thus separate what consumers must provide from what the project itself needs for testing.

\subsubsection{Oscillation across Roles}\label{prac:oscillation}
Beyond single-direction reassignment, some dependencies undergo role oscillation, where declared dependency roles may be revised multiple times.
As shown in Table~\ref{tab:practice_distribution}, oscillation appears consistently across roles, affecting 7.5\,\% of projects with Core dependencies, 9.1\,\% for Dev, and 5.2\,\% for Peer.
In these cases, role reassignment does not end with a different terminal role, but eventually returns to the starting role after one or more intermediate changes.

One situation is a rapid rollback after a reassignment cannot be sustained.
For example, \texttt{QuantiModo/quantimodo-android-chrome-ios-web-app}\footnote{\href{https://github.com/QuantiModo/quantimodo-android-chrome-ios-web-app}{https://github.com/QuantiModo/quantimodo-android-chrome-ios-web-app}} is a configurable app for tracking personal outcomes across Android, Chrome, iOS, and web.
In pull request~\href{https://github.com/QuantiModo/quantimodo-android-chrome-ios-web-app/pull/4141}{\#4141} (commit~\href{https://github.com/QuantiModo/quantimodo-android-chrome-ios-web-app/commit/e5d2e6d}{\texttt{e5d2e6d}}, 2020-10-09, message: \texttt{Feature/start}), 41 packages (mostly the project's gulp toolchain) were moved from Core to Dev.
About six hours later, pull request~\href{https://github.com/QuantiModo/quantimodo-android-chrome-ios-web-app/pull/4158}{\#4158} (commit~\href{https://github.com/QuantiModo/quantimodo-android-chrome-ios-web-app/commit/6a289edb6}{\texttt{6a289ed}}, 2020-10-10, message: \texttt{Move gulp to dependencies}) moved them back to Core.
In this project, the build runs during deployment: Heroku first installs only Core packages, then \texttt{heroku-postbuild} runs the gulp configure step.
After the move to Dev, gulp was no longer installed in that environment, so the packages were moved back to Core within hours.

Other oscillations extend beyond a single rollback and unfold through repeated revisions over a longer period.
As discussed above, \texttt{wesbos/eslint-config-wesbos} is a shareable ESLint and Prettier configuration package.
In earlier versions (e.g., \href{https://github.com/wesbos/eslint-config-wesbos/commit/f833906}{\texttt{f833906}}, 2021-12-07, message: \texttt{add support for eslint 8}), packages such as \texttt{@babel/core}, \texttt{eslint}, \texttt{eslint-plugin-react}, and \texttt{prettier} were declared under both Dev and Peer.
The later install simplification (\href{https://github.com/wesbos/eslint-config-wesbos/commit/44001c0}{\texttt{44001c0}}, 2023-01-05) moved many of these packages into Core, while keeping \texttt{eslint}, \texttt{prettier}, and \texttt{typescript} as Peer dependencies for consumers.
Subsequent updates continued to revise these roles: \href{https://github.com/wesbos/eslint-config-wesbos/commit/62b629d}{\texttt{62b629d}} (2024-08-29, message: \texttt{Removes babel parser, updates dependencies}) declared \texttt{eslint} and \texttt{prettier} under Core as well, while still retaining them as Peer.
Across these commits, the same packages are revised repeatedly as ESLint and Prettier versions, peer requirements, and consumer installation expectations change.
For a shareable config package, maintainers must decide whether tools such as \texttt{eslint} and \texttt{prettier} are needed only while developing the config, should be installed with the package, or must be provided by consuming projects.
Because these roles overlap in practice, the maintainer keeps adjusting the declarations.

Overall, developers may continue to adjust dependency roles through repeated reclassification over time.
This motivates our timing analysis in RQ4, which examines how long dependency role declarations remain before being reclassified.
\bigskip

\begin{answerbox}
\textbf{Answer to RQ3:}
Developers revise the Core--Dev boundary in both directions: moving packages to Dev to shrink the production install set, and moving them to Core when builds or installs need those packages at production time.
They also move dependencies from Peer back to package-managed roles (Core or Dev) to shift installation responsibility back to the package. 
Role oscillation covers both short rollback after a reassignment fails under deployment, and longer repeated revision as install conventions change.
These findings underscore the ongoing maintenance efforts of developers to assign suitable roles to dependencies.
\end{answerbox}

\subsection{\RQfour}\label{rq:rq4}
RQ4 examines how long different reclassification practices unfold in project history.
For each practice, we measure the duration between a dependency's initial declaration in a role and the completion of the corresponding reclassification practice, as illustrated in \autoref{fig:practice_timing}.
Overall, reclassification is a long-term maintenance activity rather than a rapid adjustment, with practices spanning a median duration of 408 days.

One-step practices, including one-step removal and direct role reassignment, usually occur only after substantial time has passed.
One-step removal takes a median of 293 days for Core dependencies, 502 days for Dev dependencies, and 223 days for Peer dependencies.
Direct reassignment follows a similar pattern.
For example, Core-to-Dev reclassification takes a median of 178 days, while Dev-to-Core takes 201 days; other direct reassignments range from 314 days (Dev-to-Peer) to 420 days (Peer-to-Dev).

\edit{
We apply Mann--Whitney tests to compare duration distributions across roles~\citep{chinthanet2021lags,romano2006exploring}.
To measure the strength of each difference, we use Cliff's $\delta$.
Following Romano et al., we interpret $|\delta|{<}0.147$ as negligible, $0.147{\leq}|\delta|{<}0.33$ as small, $0.33{\leq}|\delta|{<}0.474$ as medium, and $|\delta|{\geq}0.474$ as large.
For one-step removal, Dev differs from both other roles with small effects (Dev and Peer $\delta{=}0.25$, Core and Dev $\delta{=}{-}0.18$; both $p{<}0.001$), while the difference between Core and Peer is negligible ($\delta{=}0.07$).
For opposite-direction reassignments, each pair unfolds on a similar time scale (Core and Dev $\delta{=}{-}0.02$, $p{=}0.002$; Dev and Peer $\delta{=}0.00$, $p{=}0.96$).
Consistent with the medians above, one-step removal takes longer for Dev than for Core or Peer, whereas reassignments of the opposite direction remain close in timing.
}

These results show that even single-step practices are rarely immediate.
Instead, dependencies often remain in their original declared roles for months before developers remove them or revise their scope.
Together with the batch removal patterns observed in RQ2 (Section~\ref{rq:rq2}), these timings suggest that many one-step changes are made only after dependencies have been used in the same role for a long time, often as part of later cleanup or restructuring.

\begin{figure}
    \centering
    \includegraphics[width=\textwidth]{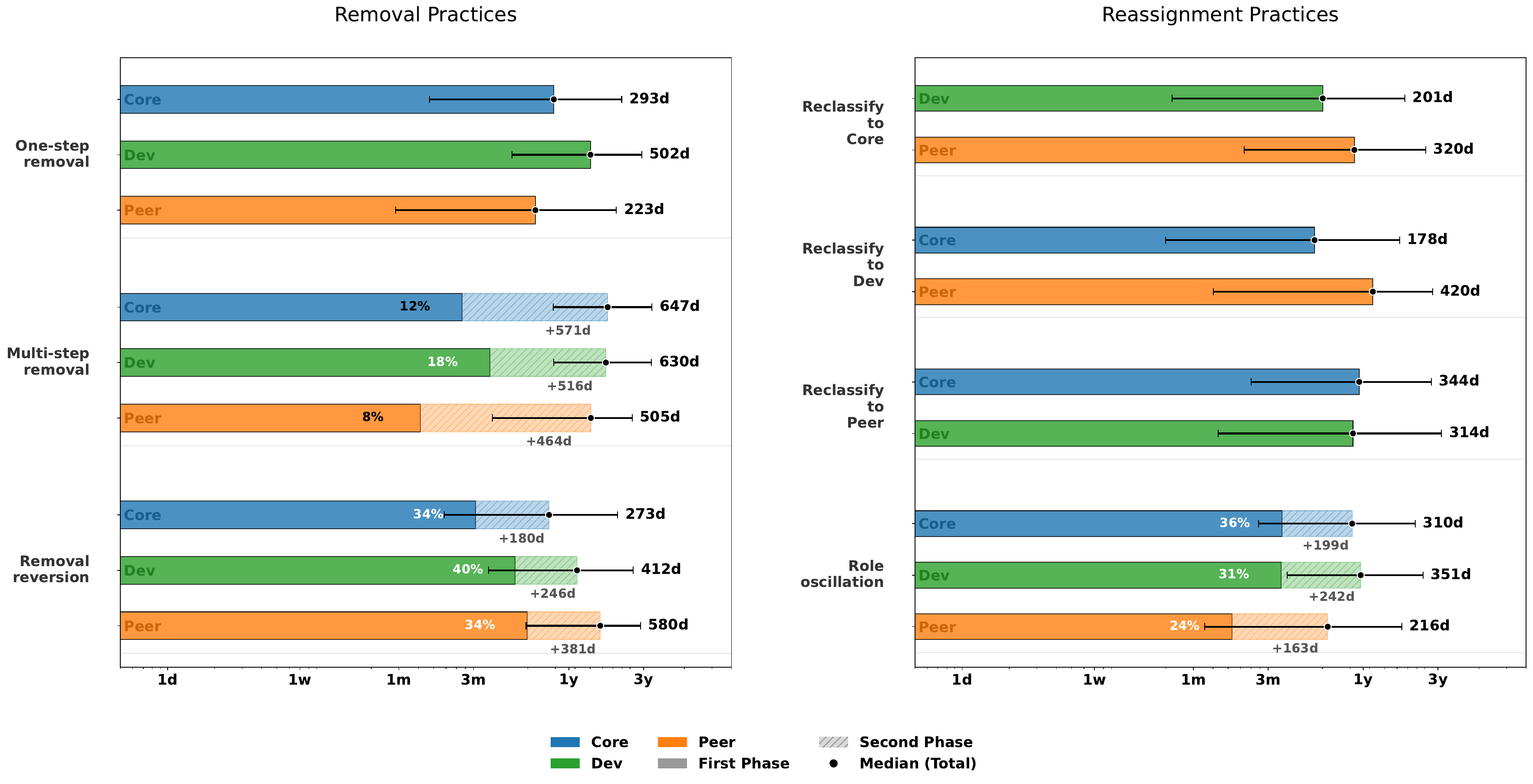}
    \caption{\scriptsize 
    Timing of dependency reclassification practices (log scale). 
    Removal (left) and reassignment (right) show median durations across roles. 
    For multi-step practices, bars are decomposed into Step 1 (time to first reclassification) and Step 2 (additional time to completion). 
    Percentages indicate Step 1's share of the total duration, while ``+Xd'' labels denote the additional median days from Step 2.
    For example, in Core multi-step removal, the total median duration is 647 days, of which Step~1 (role reassignment) accounts for only 12\,\% of the total median duration, while Step~2 (deletion) contributes an additional median of 571 days.
    }
    \label{fig:practice_timing}
\end{figure}

\bigskip
Timelines of multi-step removal follow a different pattern.
Its first reclassification occurs relatively early (within the first 20\,\% of the overall duration), especially for Core and Peer dependencies, whereas deletion happens much later.
For Peer dependencies, the first reclassification occurs after a median of 41 days and accounts for only 8\,\% of the total duration, leaving a further 464 days before deletion.
For Core dependencies, the initial Core-to-Dev/Core-to-Peer reclassification occurs after 76 days and accounts for only 12\,\% of the total duration.
In the \href{https://github.com/jenkins-infra/plugin-site}{\texttt{jenkins-infra/plugin-site}} case, the \texttt{react-dom} dependency stayed in Core for about 1,164 days before being reassigned to Dev in commit~\href{https://github.com/jenkins-infra/plugin-site/commit/2d1833a}{\texttt{2d1833a}} (2020-01-10), then remained in Dev for another 707 days before being deleted in commit~\href{https://github.com/jenkins-infra/plugin-site/commit/1031d57}{\texttt{1031d57}} (2021-12-17).
These timings suggest that in multi-step removal practices, role reassignment and deletion are temporally separated maintenance processes rather than one continuous action.

Removal reversion exhibits yet another temporal path.
Its first deletion occurs relatively early (within the first 40\,\% of the overall duration), but reintroduction takes place only after a substantial lag.
For example, for Core dependencies, deletion occurs after a median of 93 days (34\,\% of the total duration), but the reintroduction takes another 180 days.
In \href{https://github.com/infinum/eightshift-frontend-libs}{\texttt{infinum/eightshift-frontend-libs}}, the \texttt{@wordpress/server-side-render} dependency was removed from Core in commit~\href{https://github.com/infinum/eightshift-frontend-libs/commit/e66de4e}{\texttt{e66de4e}} (2020-10-08) and reintroduced to Core in commit~\href{https://github.com/infinum/eightshift-frontend-libs/commit/e6cf482}{\texttt{e6cf482}} (2024-08-07), so it stayed absent for about 1,399 days (about four years).
For Peer dependencies, the time for a reintroduction reaches a median of 381 days.
This indicates that removal reversion can involve long maintenance gaps before dependencies are restored.

Role oscillation has the shortest median duration, typically occurring within a year. 
Across roles, dependencies that start to oscillate usually experience their first role change early (within three months), followed closely (within half a year) by the next change. 
For Peer dependencies, the first reclassification occurs after a median of 53 days and the subsequent reclassification follows only 163 days later.
This suggests that when the appropriate role of a dependency is uncertain, that uncertainty tends to surface early and can trigger further role revision within a short period.
The range is still wide.
In the \texttt{QuantiModo} case (Section~\ref{prac:oscillation}), \core \arrow \develop \arrow \core\ completed in about six hours after a deploy-time build failed without Dev packages, far below the multi-month medians.
Together, these cases show what the timing distributions mean in practice: some role changes wait months or years for the next maintenance need, while others reverse within hours when deployment rejects them.

\bigskip
\begin{figure}[t]
    \centering
    \includegraphics[width=\textwidth]{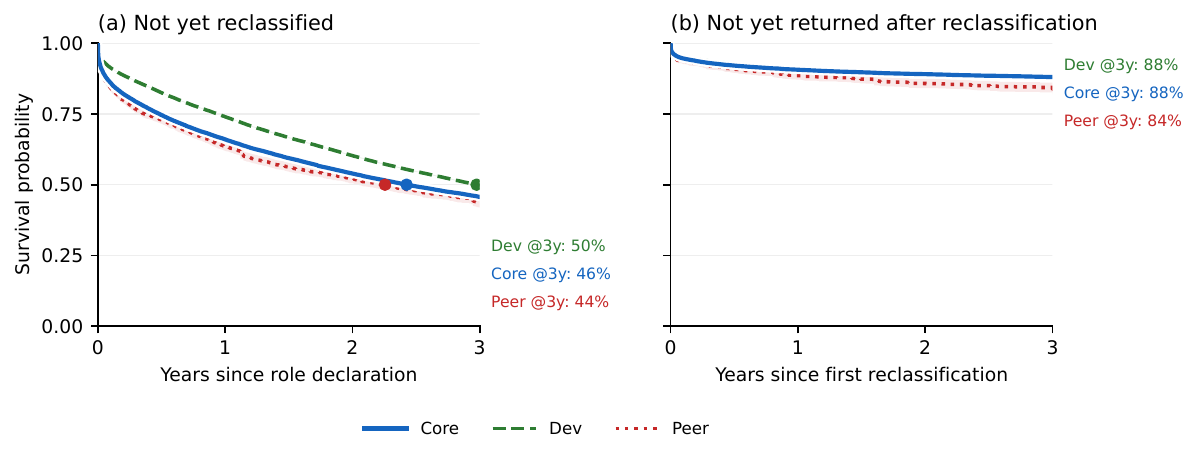}
    \caption{\small
    Time to first reclassification and return to the original role, by initial role (Core, Dev, Peer).
    (a)~Share of dependencies not yet reclassified after declaration.
    At three years: 46\,\% / 50\,\% / 44\,\% still unchanged.
    (b)~Share not yet returned to the original role after the first reclassification.
    At three years: 88\,\% / 88\,\% / 84\,\% still away (about 12\,\% / 12\,\% / 16\,\% have returned).
    }
    \label{fig:survival}
\end{figure}

The timing results above report median durations for practices that have already completed in the observed history.
They therefore omit dependencies that have not yet been reclassified by the end of the observation window, and they treat removals that have not yet been restored as if that outcome were settled.
To account for these incomplete cases, we additionally estimate how long dependencies remain in a role before a first reclassification, and how often a reclassification is later undone, while right-censoring observations that have not experienced the corresponding event by the end of each project's history (Figure~\ref{fig:survival}).

Figure~\ref{fig:survival}a shows the share of dependencies that are not yet reclassified after initial declaration.
After three years, at least half of the dependencies in every role have been reclassified at least once (54\,\% of Core, 50\,\% of Dev, and 56\,\% of Peer).
Reclassification is thus an expected stage in a dependency's life cycle, and the role assigned at introduction usually does not remain the final state.
Role review should therefore be planned as a recurring part of dependency maintenance, alongside version updates.

Figure~\ref{fig:survival}b starts from that first reclassification and shows the share that have not yet returned to the original role.
After three years, only an estimated 12\,\% of Core and Dev and 16\,\% of Peer first reclassifications have returned to the original role.
Most reclassifications are therefore durable: once a role is revised, that change usually stands.
\edit{To compare these return curves across roles, we apply log-rank tests, following prior work~\citep{decan2017depissues}.
Core and Dev follow a similar return timing ($p{=}0.12$), while Peer returns earlier than both ($p{<}0.001$), which aligns with the higher three-year return share for Peer.
}
At the same time, a non-negligible minority is later undone, so reclassification remains an iterative correction process that sometimes needs rollback.

\bigskip

\begin{answerbox}
\textbf{Answer to RQ4:}
Developers usually take months to reclassify dependencies, with a median duration of 408 days.
One-step reclassification practices typically occur after extended use (median 223--502 days across one-step removals and median 178--420 days across one-step reassignments), reflecting deliberate cleanups or adjustments.
By contrast, multi-step practices more often exhibit earlier reclassification (41--110 days) followed by significant delayed changes (further 163--571 days).
Our timing analysis indicates that dependency reclassification is a long-term maintenance process.
Our survival analysis shows that, within three years, about half of declarations have already been reclassified at least once (46\,\% / 50\,\% / 44\,\% of Core / Dev / Peer not yet reclassified), whereas only about 12\,\% of Core and Dev first reclassifications and 16\,\% of Peer first reclassifications return to that original role.
\end{answerbox}

\section{Implications for Practice and Research}
Our results show that dependency reclassification is a recurring form of dependency maintenance.
Through removals, cross-role reassignments, and later reversions, developers repeatedly revise whether a dependency should remain in the project and under which role it should be maintained.
Our timing results further indicate that declaration issues accumulate like maintenance debt: declarations that are later revised persist for months before that revision (e.g., a median of 178 days for Core-to-Dev reassignment), such revisions are deferred and settled in batches, and are often imperfect, as 62\,\% of reverted batch removals restore only a subset of the removed dependencies.

Our findings have implications for how dependency reclassification should be supported by tools, package managers, and future research.
We derive these implications from the maintenance patterns observed at scale in project histories.
They identify opportunities for support motivated by observed practice.

\subsection{Implications for Project Maintainers}
\paragraph{Audit production declarations against commonly reclassified dependencies}
RQ3 shows that Core-to-Dev reassignment is concentrated among a recurring set of development-infrastructure packages; RQ4 further shows that such revisions often occur only after extended use in the originally declared role.
Until such a revision happens, projects may retain broader production declarations than they ultimately need.
Checking the \texttt{dependencies} field against the list of frequently reclassified packages from our dataset offers a lightweight audit that can help maintainers prioritize role review.

\paragraph{Assign specific review for batch removals}
RQ2 shows that batch removal is common and that reverted batch removals often restore only part of what was deleted; RQ4 further shows that reintroduction can occur long after the removal.
Breakage caused by a batch removal may therefore remain hidden until maintainers realize that a deleted dependency is still required.
Large dependency cleanups should be treated as high-risk manifest changes, especially when dependencies serving different purposes are removed in the same commit.
Given that partial recovery of batch removal is common, isolating a large cleanup in its own change makes review and rollback easier when a dependency is still needed. 
Verifying the project after a production install (i.e., omitting \texttt{devDependencies}) can reveal missing packages before release.

\paragraph{Interpret large Dev-to-Core movements as a deployment signal}
RQ3 shows that Dev-to-Core reassignment often moves build-related dependencies into the production set.
In the cases we examined, such moves occurred when deployment required tools that had previously been declared as Dev-only.
When many build tools are moved to Core together, the change may reflect how the project is built and deployed, not only a settled decision about dependency roles.
Separating deploy and runtime stages in the pipeline may meet the same need without expanding the production install set.

\subsection{Implications for Dependency Tooling}
\paragraph{Support role review beyond version updates}
Existing dependency bots such as Dependabot\footnote{\href{https://docs.github.com/en/code-security/dependabot}{Dependabot}} and Renovate\footnote{\href{https://docs.renovatebot.com/}{Renovate}} primarily support version updates, whereas role changes receive less automated support.
However, reclassification appears in 19.4\,\% of all dependency-maintenance commits in our dataset (RQ1), indicating that role review is also a substantial part of dependency maintenance.
Tools could help identify dependencies that may warrant such review, for example by flagging development dependencies declared as Core and presenting evidence about their use in production code.
In our data, 9.8\,\% of Core dependencies are later moved to Dev, and these reclassifications are concentrated among a recurring set of development packages.
These observations can help prioritize review candidates, although the appropriate role must still be assessed in the context of each project.

\paragraph{Provide visibility into role evolution over time}
Current configuration files such as \package only record the declared role of a dependency at a specific point in time, without revealing how that role has changed across project history.
Our findings in RQ4 show that developers revise dependency roles with different maintenance timing and needs: some role changes are revisited after short intervals, whereas others remain unchanged for long periods before later adjustment.

These observations suggest that tooling should support role review from two complementary perspectives.
Within an individual project, it could provide historical context, such as when a dependency last changed roles, how long it has remained in its current role, and whether it has been reassigned multiple times.
Across projects, it could draw on maintenance data collected from many other related projects to show whether the same dependency is commonly placed under another role or frequently reclassified elsewhere.
Taken together, these two perspectives would help surface dependencies whose current role placement may require closer review.

\paragraph{Assess the impact of batch removals}
Our results in RQ2 show that dependency removals are frequently performed in batches.
When many dependencies are removed together, the challenge lies not only in deciding what to remove, but also in validating the cleanup.
This challenge is further reflected in our observations of reversal: when batch removals are later revisited, just a subset of the removed dependencies may be restored.
These reversals indicate that maintainers lack reliable ways to verify the impact of batch removals before such changes are released, which is a concrete opportunity for tool support.
For example, tooling could combine static analysis to detect remaining references, dynamic analysis to observe runtime access under representative workloads, and simulated removal in an isolated environment followed by build or test execution.

\subsection{Implications for Package Managers}
Package managers such as \npm make declared dependency roles consequential in practice by determining how dependencies are installed and who is expected to provide them.
As a result, later role reclassification is not only a change in declaration, but also a change in the practical boundary between project-managed and externally managed dependencies.

\paragraph{Warn about development-oriented dependencies installed in production scope}
Our results in RQ3 show that development-only dependencies can be used in production.
Our findings in RQ4 extend this concern by showing that related corrections often occur only after extended use, meaning that unnecessary production installation may persist for substantial periods before being revised.
The implication is therefore not limited to installation correctness at a single point in time, but also concerns inflated runtime environments and the dependency bloat they may introduce \citep{npm_dev_claim, liu2025detecting}.

These findings suggest that package managers should provide clear guidance on where dependencies should be installed via the \texttt{npm install} command, considering project context.
For example, when a dependency identified as being associated with build, test, or linting workflows is added under production dependencies (Core), the package manager should flag the assignment as unusual and prompt additional review.
Such checks are feasible as advisory warnings at install time, for example, by matching newly added Core packages against known development-oriented package families or cross-project role statistics.
Such support could make misaligned production assignments visible when they are introduced, rather than allowing them to persist until much later correction.

\paragraph{Guide Peer declarations, not only peer versions}
As shown in RQ3, Peer dependencies are often later reassigned to project-managed roles, especially Core.
In the cases we examined, such moves are sometimes motivated by simplifying the install step for downstream projects.
Package managers currently focus on missing or conflicting peer versions, but provide limited support for deciding whether a dependency should be declared as Peer in the first place.
For dependencies associated with a known host package, such guidance could consider both compatibility with the host versions used by downstream projects and whether installation should remain the consumer's responsibility.
Cross-project role histories could provide additional evidence by showing how the same dependency is declared under similar host-version settings.
For example, npm could note when a dependency is frequently reassigned away from Peer in other projects and prompt maintainers to review the choice.

\subsection{Implications for Research}
Our results from RQ3 and RQ4 suggest that future research should pay closer attention to how developers understand, assign, and later revise dependency roles.
RQ3 shows that reclassification between Core and Dev frequently occurs in both directions, especially for build tools and related infrastructure, suggesting that the boundary between runtime and development declarations is not always clear.
A later Core-to-Dev reclassification may correct an earlier misplacement, or a role that fit an earlier project setup may become misaligned only as usage, build pipelines, or packaging strategy evolve.
RQ4 further shows that such revisions often unfold over long periods of project history rather than as immediate corrections.

Together, these findings point to an important open question for research: how to decide whether a dependency's declared role still matches its current use in a project.
A crucial aspect of this research is to define clear criteria for role suitability.
This could involve statistical modeling, repository-scale empirical analysis, static analysis, and dynamic analysis to evaluate whether a dependency role is appropriate for a specific project.
While quantitative history analysis can show that these repeated revisions occur, it cannot fully explain the reasoning behind them. 
Complementary qualitative analysis of commit messages, pull requests, and issue discussions would help to explain how role decisions are made and revised as projects evolve.

\section{Threats to Validity}
\subsection{Internal Validity}
A key threat to internal validity is that our analysis is based on reclassification practices observable in the revision histories available at the time of data collection.
As a result, some practices may later evolve into other forms beyond what we observe here; for example, a removal that currently appears final may later be followed by reintroduction.
This limitation is inherent to any historical analysis of ongoing projects.
To mitigate this threat, we analyze a large and diverse corpus of \javascript projects with extensive commit histories, thereby ensuring that the practices we observe are representative of common practices across projects.
We further use survival analysis with right censoring at the end of each project's observed history, so estimates of first reclassification and return also include dependencies that have not yet experienced the event, instead of covering only changes that are already complete.

\subsection{External Validity}
Our findings are derived from the \javascript/\npm ecosystem, and their generalization depends on how other ecosystems declare dependency roles.
The distinction between runtime (Core) and development (Dev) exists in many package managers: Maven scopes (\texttt{compile}, \texttt{runtime}, \texttt{test}), Cargo's \texttt{dev-dependencies}, Composer's \texttt{require-dev}, Bundler's development groups, and Python's optional and development dependency groups all separate production requirements from development tooling.
Reclassification between such roles is expressible in these ecosystems as the same kind of manifest edit that we study, and the risk of development tools being declared in production scope is not unique to \npm.
By contrast, ecosystems that declare all dependencies in a single list, such as Go modules, cannot exhibit role reclassification.

A further consideration is that \javascript is an interpreted language whose packages are distributed as source code: \npm installs package contents directly into the project tree, where they can be loaded and executed at runtime.
A role change in \npm therefore determines which source code is physically present in a production install, making misdeclarations immediately consequential and tool support for detecting them particularly important.
The operational meaning of a role change nevertheless differs across ecosystems.
In Maven and Gradle, scopes primarily govern build-time classpaths, and in Cargo, \texttt{dev-dependencies} affect only tests and related build targets, so the same structural change may carry different consequences and different pressure to correct misdeclarations.
Moreover, \npm's Peer role has no exact counterpart: Maven's \texttt{provided} scope and Gradle's \texttt{compileOnly} configuration also delegate provisioning to the environment, but they do not declare a version-constrained requirement on a host package as Peer declarations do.
Our Peer-related findings should therefore be treated as \npm-specific.

Accordingly, we expect the structural forms we identify, which include removal, reversion, reassignment, and oscillation, to be observable in any ecosystem with role declarations, since our taxonomy is defined over transitions between declared roles.
The reported prevalence and timing, however, are \npm-specific evidence and should not be assumed to transfer; replicating the study in other ecosystems remains future work.

\subsection{Construct Validity}
Our analysis reconstructs reclassification practices from changes to the \package file on the default branch of each project.
This design captures the main development history, but it may miss changes made on long-lived branches or forks that are never merged.
In addition, a single commit may combine multiple unrelated dependency edits, while a single reclassification practice may be distributed across several commits.
These factors can obscure the local context of individual edits.
We mitigate this threat by analyzing practices at scale across more than 33K projects, which reduces the noise from fine-grained edits.
Moreover, our study analyzes declared dependency roles in the configuration file (\package), so a reclassification reflects a change in declaration context, not necessarily a complete change in how the dependency is used in the codebase.

\section{Related Work}
Prior research has studied how dependencies are maintained, how they evolve over time, and how declared dependencies relate to their actual use.
We organize related work around these three themes.
We focus on \javascript/\npm research where it is directly relevant, and include studies from other ecosystems when they address dependency bloat and role assignment.

\subsection{Dependency Management Practices and Tooling in \javascript}
Dependency management has been studied extensively across \javascript ecosystems.
Mens and Decan~\citep{mens2024overview} provide an overview of dependency challenges in package registries, including versioning, transitive dependencies, and abandonment.
A large body of work focuses on version updates and related practices:
technical lag~\citep{decan2018evolution,zerouali2018empirical},
dependency downgrades~\citep{cogo2019empirical},
update strategies and package characteristics~\citep{javan2023dependency},
semantic versioning~\citep{pinckney2023large},
library update behavior~\citep{kula2018developers,salza2018developers},
and the timing and cost of updates~\citep{berhe2020software,berhe2023maintenance,jayasuriya2022towards}.
Liu et al.~\citep{liu2022demystifying} study vulnerability propagation in \npm, and Jafari et al.~\citep{jafari2023dependency} examine dependency practices for vulnerability mitigation.

Another line of work studies tooling and automation for dependency maintenance.
In practice, bots such as Dependabot and Renovate open pull requests that update version constraints or remediate known vulnerable dependency versions.
Empirical studies of Dependabot show that such bots can reduce technical lag~\citep{he2023automating}, that many security update pull requests are accepted~\citep{alfadel2021use}, and that developers often merge bot fixes or otherwise address the corresponding vulnerabilities~\citep{mohayeji2023investigating}.
Brown and Parnin~\citep{brown2019sorry} study recommendation bots, while Kong et al.~\citep{kong2024towards} examine how developers comprehend breaking dependency changes in \npm.
Overall, current dependency bots mainly automate version updates and security upgrades.
They do not support reviewing whether a dependency should remain under specific roles, nor do they track role changes over project history.

A further group of studies concerns whether dependencies should continue to be used.
This includes alternatives for declining packages~\citep{mujahid2023go},
responses to dependency abandonment~\citep{miller2025understanding},
and characteristics of highly selected packages~\citep{mujahid2023characteristics}.

These studies cover version maintenance, automated update support, and usage decisions.
Our study examines a related maintenance activity: revising the declared role of an already introduced dependency in \package, including removal from the manifest.

\subsection{Longitudinal Dependency Evolution in \javascript and \npm}
Several studies analyze dependency evolution at the ecosystem level.
Wittern et al.~\citep{wittern2016look} study the dynamics of the \javascript package ecosystem,
Kikas et al.~\citep{kikas2017structure} analyze the structure and evolution of package dependency networks,
and Decan et al.~\citep{decan2019empirical} compare dependency network evolution across seven packaging ecosystems, including \npm.
Other work examines related ecosystem phenomena such as hidden code clones in \npm~\citep{wyss2022fork}.

Other studies examine evolution at the project level.
Terzi et al.~\citep{terzi2022software} analyze software reuse and dependency change across releases of \javascript applications.
% Ebrahimi-Kia et al.~\citep{ebrahimi2025evolution} study the evolution of direct dependencies in \npm packages.
Zerouali et al.~\citep{zerouali2026comprehensive} analyze the lifecycle of dormant \npm packages and the reactions of their dependents.
Additional studies investigate dependency adoption and minimization practices, including trivial packages~\citep{abdalkareem2017developers} and self-contained libraries~\citep{jaisri2025preliminary}.

These studies cover ecosystem growth, network evolution, dependency persistence, addition, and removal.
Our study also takes a longitudinal perspective, but follows each dependency's declared role history rather than only its presence or count.
Removal is then one possible role transition, alongside role reassignments such as Core to Dev.

\subsection{Dependency Bloat and Role Assignment}
A closely related body of work studies mismatches between dependency declarations and actual use.
In \javascript/\npm, this includes dependency smells~\citep{jafari2021dependency},
production versus declared dependencies~\citep{latendresse2022not},
unused dependencies and continuous integration waste~\citep{weeraddana2024dependency},
attack surface reduction~\citep{koishybayev2020mininode},
and bloated CommonJS dependencies~\citep{liu2025detecting}.
Wang et al.~\citep{wang2025understanding} study peer dependency resolution,
and Chuang et al.~\citep{chuang2022removing} examine the risks of removing dependencies from large projects.

Other studies develop techniques for detecting or removing unused dependency code,
including debloating approaches for JavaScript~\citep{turcotte2022stubbifier},
Java~\citep{song2024efficiently,heo2018effective,ponta2021used},
and related settings.
Beyond \javascript, dependency bloat and smells have been studied in Maven~\citep{soto2021comprehensive,soto2021longitudinal}
and Python ecosystems~\citep{drosos2024bloat,cao2022towards}.

These studies cover declaration suitability, unused dependencies, debloating, and related installation issues. 
Our study complements this literature by following these declaration changes
over project history.

\section{Conclusion}
This study presents the first large-scale longitudinal analysis of dependency reclassification across \totalprojects \javascript projects.
Our results show that dependency reclassification is a recurring part of dependency maintenance: 79.1\,\% of dependency-maintaining projects revise at least one dependency declaration after introduction.
Dependencies are frequently removed and reassigned across Core, Dev, and Peer roles, with some of these changes being reversed later.
These reclassifications often unfold through multi-stage histories and may remain separated by substantial time gaps.

Taken together, our findings show that dependency declarations cannot be treated as fixed once assigned.
A dependency's current role captures only its latest observed state, not the maintenance history that led to it.
This broadens the empirical understanding of dependency management beyond additions and version updates by showing that dependency roles themselves are maintained over time.

Our results underscore the need for role-aware support in dependency tooling and package managers, especially for tracking role histories, reviewing unstable assignments, and assessing the impact of large reclassification changes.
In future work, we plan to conduct qualitative analyses to better understand the contextual factors behind reclassification decisions, and to measure the practical consequences of reclassification (e.g., effects on install footprint, build reliability, and security exposure), in order to inform more effective guidance for dependency role maintenance.

\section*{Declarations}
\subsection*{Funding}
This work was supported by Vetenskapsrådet (Swedish Research Council) and the Knut och Alice Wallenbergs Stiftelse, with Benoit Baudry as the grant recipient.
\subsection*{Ethical Approval}
Not applicable.
\subsection*{Informed Consent}
Not applicable.
\subsection*{Author Contributions}
Yuxin Liu is responsible for the research methodology, software implementation, data collection and curation, formal analysis, visualization, as well as writing and editing the manuscript.
Cristian Bogdan provides supervision and contributes to reviewing and editing the manuscript. 
Benoit Baudry contributes to the conceptualization of the research, provides supervision, contributes to reviewing and editing the manuscript, and secures the funding.
\subsection*{Data Availability Statement}
The data and replication materials that support the findings of this study are available at:
\url{https://github.com/ASSERT-KTH/dependency-reclassification}
\subsection*{Conflict of Interest}
The authors declare that they have no conflict of interest.
\subsection*{Clinical Trial Number}
Not applicable.

\bibliography{main}

\end{document}